\documentclass[fleqn,usenatbib]{mnras}
\usepackage{fix-cm}
\usepackage{newtxtext,newtxmath}
\usepackage[T1]{fontenc}
\DeclareRobustCommand{\VAN}[3]{#2}
\let\VANthebibliography\thebibliography
\def\thebibliography{\DeclareRobustCommand{\VAN}[3]{##3}\VANthebibliography}

\usepackage{longtable}
\usepackage{booktabs}
\usepackage{bm}
\usepackage{physics}
\usepackage{graphicx}
\usepackage{color}
\usepackage{natbib}
\usepackage{comment}
\usepackage{xcolor}

\newcommand{\lya}{Ly$\alpha$}
\newcommand{\HI}{\mbox{H\,{\sc i}}}

\newcommand{\atonhe}{\textsc{aton-he}}
\newcommand{\pgt}{\textsc{p-gadget3}}

\title[LAE Bispectrum]{The Lyman-$\alpha$ emitter bispectrum as a probe of reionization morphology}

\author[Maitra et al.]{
Soumak Maitra$^{1}$\thanks{E-mail: soumak.maitra@theory.tifr.res.in},
Girish Kulkarni$^{1}$,
Shikhar Asthana$^{2}$,
James S. Bolton$^{3}$,
Martin G. Haehnelt$^{2}$ and
\newauthor\mbox{} Laura Keating$^{4}$ \\
\\
$^{1}$Tata Institute of Fundamental Research, Homi Bhabha Road, Mumbai 400005, India\\
$^{2}$Kavli Institute for Cosmology and Institute of Astronomy, Madingley Road, Cambridge, CB3 0HA, UK\\
$^{3}$School of Physics and Astronomy, University of Nottingham, University Park, Nottingham, NG7 2RD, UK\\
$^{4}$Institute for Astronomy, University of Edinburgh, Blackford Hill, Edinburgh, EH9 3HJ, UK
}

\date{Accepted ---. Received ---; in original form ---}

\pubyear{\the\year{}}

\begin{document}
\label{firstpage}
\pagerange{\pageref{firstpage}--\pageref{lastpage}}
\maketitle

\begin{abstract}
  Ly$\alpha$ emitters (LAEs) have now been discovered out to redshift $z=13$, and are valuable probes of the reionization history at redshifts beyond the reach of other currently available tracers. Most inferences of the neutral hydrogen fraction from LAE observations rely on one-point and two-point statistics like the luminosity function and the power spectrum.We present here an analysis of the bispectrum of high-redshift LAEs and demonstrate its sensitivity to the Epoch of Reionization. We use the Sherwood-Relics suite of cosmological hydrodynamical simulations post-processed with the GPU-based radiative  transfer code \atonhe\ to  generate realistic LAE mock catalogues for  a wide range of reionization models,  varying  the ionization history and the  source populations, including  contributions of AGN to hydrogen reionization. We demonstrate that the bispectrum provides greater sensitivity than the power spectrum to both the timing and spatial morphology of reionization. Using reduced-$\chi^2$ analysis we further show that the bispectrum also responds more strongly to variations in source population and AGN contribution, apparently more efficiently capturing morphological features missed by two-point statistics. The redshift evolution of the bispectrum reflects the increased clustering of ionizing sources at earlier epochs.  The sensitivity of the bispectrum to peculiar velocities underscores the importance of velocity corrections in comparisons to observations.   Our findings  demonstrate that the  LAE bispectrum is a powerful higher-order statistic for probing reionization through current and future LAE surveys using telescopes such as Subaru and JWST.
\end{abstract}




\section{Introduction}

\begin{table*}
\centering
\caption{Summary of observational programs targeting the three-dimensional distribution of LAEs. We include narrow-band, spectroscopic, and IFU-based surveys spanning a wide range of redshifts and volumes. Future surveys such as Roman will greatly expand the accessible 3D LAE sample.}
\label{tab:lae_surveys}
\begin{tabular}{lccccc}
\hline
\textbf{Survey} & \textbf{Instrument} & \textbf{Redshift Range} & \textbf{Area (deg$^2$)} & \textbf{3D Capability} & \textbf{Type} \\
\hline
SILVERRUSH     & Subaru/HSC (NB)          & 2.2–7.3     & $\sim$25    & Partial (NB slices)     & Ground (Completed) \\
PFS-SSP    & Subaru/PFS    & 5.7-6.6     & $\sim$15    & Full 3D (spectra)                & Ground (Ongoing) \\
DAWN           & NEWFIRM (NB)             & 7.7       & $\sim$1   & Partial (NB)            & Ground (Completed) \\
MUSE-Wide      & VLT/MUSE (IFU)           & 2.9–6.7   & $\sim$0.03  & Full 3D                 & Ground (Completed) \\
MUSE-HUDF      & VLT/MUSE (IFU)           & 2.9–6.7   & $\sim$0.003 & Full 3D                 & Ground (Completed) \\
LAGER          & DECam (NB964)            & 6.9       & $\sim$3   & Partial (NB)            & Ground (Ongoing) \\
CIDER          & NB964/NB1008             & 6–7.5     & -     & Partial (NB)            & Ground (Ongoing) \\
WERLS          & Keck/MOSFIRE+LRIS        & 6–8       & $\sim$0.7   & Full 3D (spectra)       & Ground (Ongoing) \\
MOONS          & VLT/MOONS (Multi-object) & 6–8.5     & $\sim$1–2   & Full 3D (spectra)       & Ground (Future) \\
CEERS          & JWST/NIRSpec             & 6–9       & $\sim$0.03  & Full 3D (spectra)       & Space (Ongoing) \\
JADES          & JWST/NIRSpec+NIRCam      & 5.8–8.0   & $\sim$0.05  & Full 3D                 & Space (Ongoing) \\
COSMOS-Web     & JWST/NIRCam              & 6–11      & 0.54        & Imaging + WERLS overlap & Space (Ongoing) \\
Roman Grism    & Roman Space Telescope    & 6–10+     & $\sim$16    & Full 3D                 & Space (Future) \\
\hline
\end{tabular}
\end{table*}

The Epoch of Reionization (EoR) represents a critical phase transition in cosmic history, during which ultraviolet (UV) photons emitted by the first stars and galaxies ionized the neutral hydrogen in the intergalactic medium (IGM). While early observations of Gunn–Peterson troughs in the spectra of $z \gtrsim 6$ quasars indicated that the IGM remained substantially neutral until $z \sim 6$ \citep{Fan2006, Becker2015}, more recent studies suggest that reionization likely ended later, around $z \sim 5.3$ \citep{kulkarni2019, keating2020,Bosman2022}. These observations imply a late end to reionization, characterized by a rapid rise in the ionized hydrogen fraction over a short interval in cosmic time.
Independent constraints from polarization anisotropies in the cosmic microwave background (CMB), including results from the \textit{Planck} satellite \citep{planck2018cp} and subsequent analyses \citep{deBelsunce2021, Giare2024}, provide integrated Thomson scattering optical depths consistent with reionization midpoints typically around \( z \sim 7\text{--}8 \), although exact values vary depending on assumptions about the reionization history. 
Evidence from quasar absorption spectra further supports a patchy and extended reionization process. Studies of Ly$\alpha$ damping wings in high-redshift quasars provide model-dependent constraints on the volume-averaged neutral hydrogen fraction during the Epoch of Reionization, with several works reporting a significant neutral fraction of $\bar{x}_{\mathrm{HI}} \sim 0.5$ at $z \sim 7$ \citep{davies2018, Satyavolu2023a, Satyavolu2023b, Greig2024, Durovcikova2024, Becker2024, Sawyer2025}. Complementary, model-independent upper limits have been obtained using the dark-pixel fraction method in the Ly$\alpha$ and Ly$\beta$ forests \citep{McGreer2015, jin2023}.
Deep imaging and spectroscopic campaigns have also constrained the evolution of galaxy populations responsible for reionization. Measurements of the UV luminosity function at $z \gtrsim 6$ suggest that faint galaxies dominate the ionizing photon budget, with a steepening faint-end slope and rapidly evolving number density \citep{Bouwens2015, Liu2016, Harikane2022}. While these studies do not directly constrain the escape fraction of ionizing photons ($f_{\mathrm{esc}}$), their implications for reionization require assumptions about the fraction of ionizing photons that escape from galaxies into the intergalactic medium. Recent studies by \citet{Simmonds2023, Simmonds2024} demonstrate that low-mass, bursty galaxies possess high ionizing photon production efficiencies ($\xi_{\mathrm{ion}}$) and, with modest escape fractions (10–20\%), could account for the ionizing photon budget required for reionization. More recently, the detection of Ly$\alpha$ emission from a galaxy at $z = 13$ in the JWST JADES survey indicates that ionized bubbles may have already formed as early as 330 Myr after the Big Bang \citep{Witstok2025}. 
This suggests that reionization activity—and therefore non-zero ionized hydrogen fractions—may already be present at $z > 10$, pointing to a much earlier onset of reionization than previously assumed \citep{asthana2024}.  While galaxies are considered the principal drivers of reionization, the contribution from faint active galactic nuclei (AGNs) and obscured quasars remains under active investigation \citep{Matsuoka2018, Dayal2018,asthana2024b}.

While many works have focused on constraining the \textit{mean} neutral fraction ($\bar{x}_\mathrm{HI}$) at different epochs \citep{Planck2016re,mason2018,mason2019,hoag2019,morales2021,jin2023}, far less is known about the \textit{spatial distribution} of neutral hydrogen, or equivalently, the patchiness and morphology of reionized bubbles \citep{Becker2021Morphology,Becker2021BubbleStats, Christenson2021, Gangolli2024}. The spatial morphology of reionization is predicted to be highly inhomogeneous, with ionized regions percolating outwards from clustered galaxies and merging over time \citep{furlanetto2004, Sobacchi2014, Mesinger2016, Naidu2020}.  A detailed understanding of the 
spatial distribution of the neutral fraction will, thus, be pivotal for identifying the sources driving reionization \citep{robertson2022,mascia2023} and for understanding the feedback of reionization on galaxy formation \citep{simpson2013,geil2016,furlanetto2017}.
 Moreover, knowing how the ionized and neutral regions are distributed in the IGM can provide excellent priors for future 21\,cm studies, which suffer from intense foreground contamination at low radio frequencies. For instance, morphological information extracted from galaxy surveys (e.g., the typical size of ionized bubbles) could inform or break degeneracies in 21\,cm analyses \citep[e.g.,][]{McQuinn2007, Mesinger2011, Ghara2020}.

High-redshift LAEs, which are star-forming galaxies with strong Ly$\alpha$ emission, have emerged as key probes of the reionization era, uniquely enabling the study of the ionization state of the IGM at redshifts $z \gtrsim 6$ where other direct observables are limited. Unlike 21\,cm experiments, which remain in the early stages due to foreground contamination and calibration challenges, LAEs are currently observable with existing ground- and space-based instruments, providing crucial insights into the topology and timeline of reionization well before large-scale 21\,cm tomography becomes feasible. Ly$\alpha$ photons resonantly scatter off neutral hydrogen, so the visibility of LAEs is sensitive to the neutral hydrogen fraction of the ambient CGM and IGM \citep{Dijkstra2014, Sadoun2017}. If the hydrogen in the IGM is still significantly neutral, Ly$\alpha$ emission is heavily attenuated by damping-wing absorption, whereas in ionized regions Ly$\alpha$ photons can redshift out of resonance in the ISM of the emitting galaxy and evade absorption by the intervening IGM. Thus, the observed abundance and clustering of LAEs at $z \gtrsim 6$ carry imprints of how reionization proceeds. Narrowband surveys such as SILVERRUSH (Subaru/HSC) find that the LAE luminosity function (LF) declines from $z \sim 5.7$ to $z \sim 7$ \citep{konno2014, Ouchi2018}, suggesting a rising neutral fraction that still heavily attenuates bright LAEs at $z \sim 7$. Indeed, the abundance of LAEs plummets beyond $z \sim 7$, consistent with late reionization where a significant neutral fraction ($x_{\mathrm{HI}} \gtrsim 0.5$) persists at $z \gtrsim 7$ \citep{Ota2017, mason2019, Jung2020}. Complementary studies of Lyman-break galaxies also find a declining fraction of Ly$\alpha$-emitting galaxies from $z \sim 6$ to 8, reinforcing the picture of an increasingly neutral IGM at $z \sim 7$–8 \citep{Pentericci2014, Stark2017, Endsley2022, Chen2024jwst}. Intriguingly, recent JWST/NIRSpec observations have detected Ly$\alpha$ emission in galaxies at $z > 8$ \citep{ Tang2023CEERS,Nakane2024}, indicating early ionized bubbles likely driven by local galaxy overdensities. Taken together, these probes suggest we are witnessing the tail end of reionization at $z \sim 6$–7, although degeneracies remain between the evolution of the IGM and that of the CGM and the intrinsic properties of galaxies \citep{Weinberger2018, Rhoads2003, Malhotra2004}.

The spatial distribution of LAEs during the EoR is anticipated to closely reflect the intrinsically patchy morphology of cosmic reionization \citep{McQuinn2007, Iliev2008, Mesinger2011, Jensen2013, Hutter2015, Weinberger2018}.
 Ionized bubbles, formed around clustered LAEs and other luminous sources, are predicted to extend up to tens of megaparsecs (Mpc) before eventually overlapping \citep{McQuinn2007,Iliev2008,Dijkstra2011,hayes2023,lu2024,neyer2024}.
 If reionization is not yet complete, LAEs are expected to preferentially reside within large ionized regions, resulting in enhanced clustering at intermediate scales due to the neutral intergalactic medium (IGM) effectively acting as a selection filter \citep{McQuinn2007,Mesinger2011}. Consequently, two-point statistics, such as power spectra and correlation functions of high-redshift LAEs, have been widely employed to search for signatures indicative of ongoing reionization \citep{Ouchi2010,Ouchi2018}. Nonetheless, the complex topology and inherently non-Gaussian structure of reionization introduce features that two-point statistics alone cannot fully capture. This limitation motivates the exploration of higher-order statistics, notably the bispectrum---the Fourier-space counterpart to the three-point correlation function---as a more sensitive and robust probe of non-Gaussian signatures inherent to reionization \citep{Yoshiura2015,Yoshiura2017,Majumdar2018}.

The bispectrum has long been employed in other areas of cosmology (e.g., the CMB, large-scale galaxy surveys, and 21\,cm surveys), precisely because it captures mode-coupling and non-linear effects that the power spectrum alone misses. In 21\,cm cosmology, for instance, the bispectrum can directly probe the shape and size distribution of ionized bubbles during the EoR and even track ``sign flips'' as reionization progresses \citep[e.g.,][]{Majumdar2018, Ghara2020,Raste2023}. In CMB studies, bispectrum measurements constrain primordial non-Gaussianities and late-time lensing or secondary anisotropies \citep[e.g.,][]{Planck2016NG}. In large-scale structure surveys (e.g., BOSS, DESI), the galaxy bispectrum helps break degeneracies in bias parameters and growth rates \citep[e.g.,][]{GilMarin2015, Slepian2017}. Despite these diverse applications, the bispectrum has not yet received extensive observational attention for reionization-era \textit{LAEs}, primarily due to the scarcity of large 3D samples at $z \gtrsim 6$. While the potential of the LAE bispectrum to disentangle radiative transfer effects from gravitational clustering was first discussed by \citet{Greig2013}, its utility as a probe of ionized structure during the EoR remains largely unexplored.

Observationally, the advent of large LAE surveys provides the necessary data volume to attempt bispectrum measurements. Ground-based efforts have played a central role in mapping the LAE distribution at cosmic dawn. The SILVERRUSH survey \citep{Ouchi2018} has mapped thousands of LAEs at $z \sim 5.7$--$6.6$ over several square degrees using the Subaru telescope. The LAGER project \citep{Zheng2017} represents one of the most extensive narrowband LAE efforts at $z \sim 6.9$, with over 300 LAEs presented by \citet{Harish2022}, including a prominent bright-end excess in some fields and constraints on the neutral fraction $x_{\rm HI}$ from \citet{Sierralta2024}. Earlier luminosity function results include \citet{Zheng2017} and \citet{Hu2019}, while \citet{Hu2021} reported a $z = 6.9$ protocluster in the COSMOS field—an ideal target for tomographic reconstruction, especially with ongoing JWST campaigns in this region. The CIDER project (ongoing) further targets LAE-based tomography using narrowband filters NB964 and NB1008. Complementary spectroscopic efforts include the MUSE instrument on the VLT, which has pioneered 3D LAE mapping. Surveys such as MUSE-Wide and MUSE-HUDF have identified thousands of LAEs between $2.9 \lesssim z \lesssim 6.7$ \citep{Bacon2015, Bacon2017, Inami2017, Urrutia2019}. The upcoming VLT MOONS spectroscopic survey will extend this legacy to wider areas with high multiplexing, enabling deep spectroscopic observations of faint galaxies at $z > 6$ \citep{Cirasuolo2020}.

Space-based surveys have pushed LAE detections to higher redshifts and fainter luminosities. JWST campaigns—including JADES \citep{Eisenstein2023}, CEERS, COSMOS-Web \citep{Casey2023}, PRIMAL, and UNCOVER—have begun to reveal the diversity of LAEs at $6 \lesssim z \lesssim 9$. Spectroscopic and IFU-based observations are transforming our ability to construct the full three-dimensional distribution of LAEs. Unlike narrowband imaging, which probes thin redshift slices, spectroscopic surveys recover the line-of-sight dimension, enabling genuine 3D clustering analyses, including the bispectrum. Recent analyses by \citet{Jones2023} and \citet{Tang2024} reveal diverse Ly$\alpha$ line profiles and ISM properties, underscoring the complex ionization and kinematic environments of these early galaxies. The WERLS project (Keck/MOSFIRE+LRIS) targets $\sim800$ UV-selected galaxies at $6 < z < 8$, overlapping with JWST fields such as CEERS and COSMOS-Web, and probing ionized bubbles on 10–100 Mpc scales. The DAWN survey \citep{Tilvi2020}, along with earlier programs like \citet{Zitrin2015} and \citet{Larson2023}, have confirmed LAEs up to $z \sim 8.6$.

Looking forward, next-generation facilities will revolutionize the 3D mapping of LAEs. Nancy Grace Roman Space Telescope will offer nearly 100 times the field-of-view of HST or JWST, enabling wide-field 3D LAE catalogs essential for higher-order clustering statistics such as the bispectrum \citep{Yung2023}. The Subaru Prime Focus Spectrograph (PFS), with its wide field-of-view and high multiplexing capability, is poised to carry out deep spectroscopic surveys of LAEs at $z>5.7$, bridging the gap between imaging-selected samples and precise large-scale structure measurements.  Furthermore, cross-correlating JWST-selected galaxies with background QSOs in datasets such as EIGER \citep{Matthee2023EIGER} and ASPIRE \citep{Wang2023ASPIRE} offers a promising route to jointly constrain the small- and large-scale structure of neutral hydrogen. With these advances, direct tomographic reconstruction of ionized regions using galaxy distributions has become an increasingly viable goal.
A summary of current and upcoming efforts for 3D LAE mapping is provided in Table~\ref{tab:lae_surveys}.

In this paper, we investigate the potential of using the bispectrum of \lya\ emitters as a morphological tracer during the Epoch of Reionization. Utilizing the cosmological hydrodynamical Sherwood-Relics simulations \citep{puchwein2023} post-processed with the GPU-based radiative transfer code  \atonhe\ \citep{asthana2024}, we generate mock LAE populations for various reionization models. Our goal is to quantify how effectively the LAE bispectrum captures non-Gaussian morphological features, distinguishing among different reionization models and improving upon limitations inherent in two-point statistics. This paper is structured as follows: Section~\ref{Sec:Simulation} describes our simulations and reionization models. In Section~\ref{Sec:Methodology}, we detail our methodology for generating realistic LAE mock catalogs, including intrinsic luminosity assignments and intergalactic medium attenuation. Section~\ref{Sec:Clustering} presents our analysis of the three-dimensional clustering of LAEs, emphasizing comparisons between power spectrum and bispectrum results across different reionization histories. Finally, we summarize our key conclusions and discuss observational prospects with forthcoming galaxy surveys in Section~\ref{Sec:Conclusion}.

\section{Reionization Simulations}
\label{Sec:Simulation}

To model reionization, we use radiative transfer cosmological simulations performed using the \atonhe\ and \pgt\ codes.  The radiative transfer is done in post-processing: hydrodynamical simulations are first performed using \pgt, and then the radiative transfer and gas thermochemistry is computed using \atonhe. We now give these details of these simulations and the reionization models considered.

\subsection{Hydrodynamical simulation of the IGM}

We model the IGM using one of the cosmological hydrodynamical simulations from the Sherwood-Relics suite of simulations \citep{puchwein2023}.  This was run with the \pgt\ code (a modified version of {\sc Gadget-2} code described by \citealt{springel2005}), in a periodic box with volume ($160\, h^{-1}$cMpc)$^3$ with $2048^3$ gas and $2048^3$ dark matter particles. The initial conditions were evolved from $z=99$ to $z=4$ and the simulation snapshots were saved in intervals of 40~Myr. To make the simulation run faster,the \texttt{QUICK\_LYALPHA} prescription was used in  \pgt\ that converts gas particles with temperature less than $10^5$K and overdensity more than 1000 to star particles \citep{viel2004a}. The gas densities were then projected on a Cartesian grid of ${2048}^3$ cells, on which the radiative transfer is performed in post-processing using  \atonhe\ \citep{asthana2024}. To approximate the impact of photoionization heating and pressure smoothing of the gas during reionization, a uniform UV background \citep{puchwein2019} was employed in the hydrodynamical run. Note that while \citet{puchwein2023} also present hybrid simulations including radiative transfer, in this work we utilize only their hydrodynamical simulation runs that incorporate a uniform UV background.

\subsection{Reionization modelling with \atonhe}
\begin{figure*}
	\includegraphics[width=0.7\textwidth]{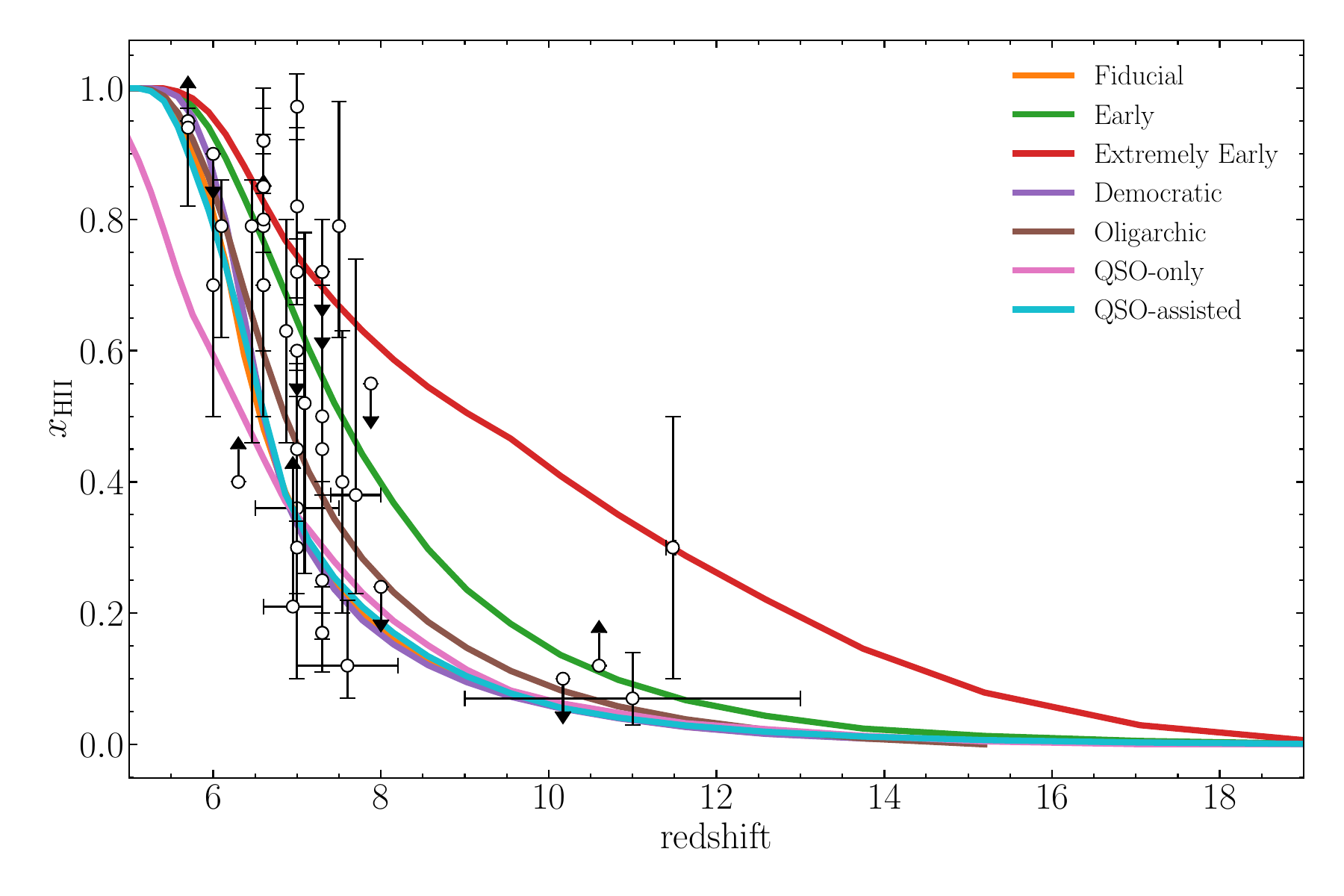}%
	\caption{
    Evolution of the volume-averaged ionized hydrogen fraction, $x_{\mathrm{HII}}$, for seven of the eight reionization models considered in this study. The eighth model, MP19, assumes a spatially uniform ultraviolet background and is therefore excluded from this figure. All models are calibrated to reproduce the observed mean \lya\ forest transmission at \( z \lesssim 6.5 \). See Table~\ref{tab:simulations} and the main text for details of the model parameters and calibration procedure. Overplotted data points show observational inferences of $x_{\mathrm{HII}}$ based on a variety of methods: measurements of the \lya\ equivalent width in high-redshift Lyman-break galaxies \citep{2015MNRAS.446..566M, hoag2019, mason2019, 2023ApJ...949L..40B, 2023ApJ...947L..24M}; LAE two-point clustering analyses \citep{Ouchi2018, 2024ApJ...971..124U}; LAE luminosity function studies \citep{Ouchi2010, konno2014, Konno2018, morales2021, 2021ApJ...923..229G, 2018PASJ...70...55I, 2024ApJ...971..124U}; and \lya\ damping wing constraints from spectra of galaxies, quasars, and gamma-ray bursts \citep{2006PASJ...58..485T, 2013MNRAS.428.3058S, 2014PASJ...66...63T, davies2018, 2019MNRAS.484.5094G, 2020ApJ...896...23W, 2023NatAs...7..622C, Durovcikova2024, 2024ApJ...973....8H, 2024ApJ...971..124U}.}
\label{Fig:xHI}
\end{figure*}

The radiative transfer was performed in post-processing using \atonhe\ \citep{asthana2024} which is a revised version of the hydrogen-only radiative transfer code \textsc{aton} \citep{aubert2008}. \atonhe\ improves upon \textsc{aton} by incorporating helium, and allowing  multi-frequency solutions of the radiative transfer equation. All of our reionization models were run in the multi-frequency mode, except for the Extremely Early model, which was executed in a single-frequency configuration. In all of our models apart from the Democratic source model the ionizing sources were modeled by assigning ionizing emissivities to the dark matter halos proportional to the mass of the halo, following \citet{kulkarni2019,keating2020}. This assumption is based on the fact that star-formation rate at these redshifts is expected to be proportional to the halo mass. A cut-off of $10^9$~M$_{\odot}h^{-1}$ was introduced and only halos more massive than this cut-off were used for source modelling
in all of our source models apart form the Oligarchic source models where a higher cut-off mass of $8.5 \times 10^9$~M$_{\odot}h^{-1}$ was assumed. The spectrum of each source was then modeled using a black-body spectrum. In this work, we use a black-body source spectrum corresponding to $T=4\times 10^4\,$K. \atonhe\ then solves the radiative transfer equation in a moment-based approximation using the M1 closure \citep{levermore1984}, while simultaneously evolving the ionization fractions and the gas temperature on the grid.
We consider multiple distinct reionization histories, all of which are extensively described in \citet{asthana2024,asthana2024b}. Each reionization model is carefully calibrated by adjusting the emissivities of ionizing sources to reproduce the observed  mean flux of the Ly$\alpha$ forest transmission at $z<6$. The details of these calibrated reionization histories and their $\left<X_{\rm HII}\right>$ evolution are illustrated in Figure~\ref{Fig:xHI}.:

\begin{itemize}
\item\textit{Late (Fiducial) Reionization History}: This model is calibrated by tuning the emissivity of ionizing sources to reproduce observed mean Ly$\alpha$ forest transmission at $z<6$, with the midpoint of hydrogen reionization at $z_{\mathrm{mid}}\sim 6.5$. Reionization completes by approximately $z\sim5.5$ in this model. Ionizing sources are massive stars in galaxies, with emissivities  proportional to their halo masses above the $10^9\, M_{\odot}h^{-1}$ threshold. In this model, the IGM experiences a gradual temperature increase due to ionizing photons from galaxies, peaking at approximately 10,000 K. Post-reionization, the IGM cools down as the ionizing sources diminish.

\item\textit{Early and Extremely Early Reionization Histories}: We also consider two models with the same source model as in the Fiducial model where hydrogen reionization occurs earlier than in the Fiducial model. The \textit{Early} model features a midpoint around $z_{\mathrm{mid}} \sim 7.5$ and, similar to the Fiducial model, is calibrated with Ly$\alpha$ forest data but assumes enhanced early ionizing emissivity. The \textit{Extremely Early} model is characterized by a significantly earlier reionization history, with a midpoint redshift $z_{\mathrm{mid}} \sim 9.5$. This model is executed in a hydrogen-only, single-frequency mode and provides a strong contrast with the Fiducial case, especially regarding thermal and ionization structures at lower redshifts. Reionization in this model reaches its midpoint at  $z \sim 8.5$. Such early reionization histories may be required to explain the high-redshift LAE detections, even though they are in tension with CMB constraints \citep{asthana2024}. The ionizing emissivity in the Extremely Early model is substantially elevated at early times and declines rapidly towards lower redshifts.
Recent CMB analyses suggest that allowing a larger optical depth can relieve several late-time tensions (e.g., $H_0$, $S_8$) but creates a new mismatch with the low-$\ell$ Planck polarization data \citep{Yeung2024, Giare2024, Sailer2025}. These works highlight both the promise and the current uncertainty in $\tau$, emphasizing the need for independent probes of reionization.

\item\textit{QSO-assisted source model}: JWST observations suggest that the number density of QSOs/AGN at $z>4$ could be higher by as much as an order of magnitude or more than previously thought, particularly at faint magnitudes \citep{onoue2023,furtak2023,harikane2023,kokorev2023,goulding2023,maiolino2023,greene2024,matthee2024,kocevski2024,lin2024,akins2024,fujimoto2024,labbe2025}. To study the impact of such sources, we include a model where faint AGN contribute significantly (around 17\%) to the hydrogen-ionizing photon budget alongside galaxies. Here, AGN luminosities and number densities are calibrated to match Ly$\alpha$ forest data, particularly at intermediate redshifts ($5 \lesssim z \lesssim 6.2$). In this hybrid model, halos with $M > 10^9\,M_\odot/h$ host both stellar sources and AGN, with 1\% of halos randomly seeded with transient AGN of 10 Myr lifetime, contributing 17\% of the total ionizing emissivity; QSOs emit using a multi-frequency power-law spectrum ($L_\nu \propto \nu^{-1.7}$), while the remaining 83\% comes from galaxies using a stellar spectrum. QSOs, with their hard spectra, significantly ionize helium, leading to earlier HeII reionization. The QSO-assisted model results in a more extended and patchy reionization process, with higher temperatures due to the additional ionizing photons from QSOs. 
 This model addresses recent observational suggestions of early quasar activity indicated by JWST-selected high-redshift Ly$\alpha$ emitting galaxies (LAEs) and is consistent with Planck optical depth measurements, provided the QSOs complement galaxy-driven reionization.

\item\textit{QSO-only source model}: A more extreme model considers reionization driven entirely by QSOs/AGN, adjusted to match mean Ly$\alpha$ forest transmission observations. This model excludes stellar sources entirely and assigns 10\% of halos as QSO hosts, each with a 10 Myr lifetime and collectively supplying 100\% of the ionizing photon budget through a power-law spectral energy distribution covering hydrogen and helium ionization energies. Due to the scarcity of QSOs and their intense ionizing output, reionization is highly inhomogeneous, with large ionized regions around QSOs and vast neutral areas elsewhere. As extensively discussed in \citet{asthana2024b}, this model tends to yield excessively high gas temperatures in the post-reionization IGM at $z<5$, potentially conflicting with observational constraints unless the escape fraction of He\,\textsc{ii}-ionizing photons is suppressed. Similar considerations have also been explored in the context of QSO-driven reionization by \citet{madau2024}.

\item \textit{Democratic source model}:
  In this model reionization is driven by numerous faint galaxies, with all halos above $10^9\,M_\odot/h$ contributing ionizing photons equally independent of their mass. Ionizing emissivity is based on stellar sources. The Democratic model assumes that both low- and high-mass galaxies contribute comparably to reionization, resulting in a more uniform and gradual ionization history with moderately heated IGM temperatures. The model is calibrated to match the \lya\ forest constraints at $z<6$.

  \item \textit{Oligarchic source model}:
  In this model, a smaller number of more massive galaxies dominate the ionizing photon budget. Only halos above an increased mass limit of $8.5 \times 10^9\,M_\odot/h$ are included,  effectively privileging more massive halos than in the fiducial source model. This results in a more patchy reionization. As with the Democratic model, stellar sources are modelled with stellar spectrum, and the model is calibrated against \lya\ forest constraints at $z<6$.

  \item\textit{Uniform UV Background Model}: 
To understand if the bispectrum is sensitive to the patchiness in the distribution of neutral hydrogen, we also include a comparison with a simplified benchmark model based on the default simulation presented in \cite{puchwein2023}, which employs a uniform UV background as described in \cite{puchwein2019}. We modify the \cite{puchwein2019} background by applying a constant re-scaling of the \HI\ photo-ionization rates so that the volume-averaged $\left<n_{\rm HI}\right>$ matches that of the Fiducial model. This model assumes photoionization equilibrium, with the gas temperature set by balancing photoheating and adiabatic/cooling losses as originally tabulated. We will refer to this rescaled uniform UVB model as the "MP19" model. While not intended as a physically realistic scenario, this toy model serves as a useful reference to isolate and quantify the effects of spatial inhomogeneities, or "patchiness", in the reionization history. By contrasting the patchy reionization models with this idealized uniform background case, we aim to elucidate how fluctuations in the ionizing radiation field impact the morphology of reionization.
\end{itemize}

In this work, we decided to work with distinct reionization models, instead of a grid of models, because all the reionization models mentioned above (except, MP19)  are tightly constrained by available observational datasets such as the JWST LAEs number counts, Planck CMB optical depth measurements, ionizing efficiencies required by the clumping factor constraints, and IGM temperature and transmission data from the Ly$\alpha$ forest. \cite{asthana2024,asthana2024b, asthana2024c} extensively evaluated the consistency of these models with observations, providing a comprehensive framework for interpreting the astrophysical sources responsible for reionization and their observable signatures.

\begin{table*}
\centering
\caption{Summary of reionization models used in this work. The quantity \( M_{\rm min} \) denotes the minimum halo mass hosting ionizing sources. The parameter \( \dot{N}_{\rm QSO}/\dot{N}_{\rm total} \) specifies the fractional contribution of QSOs to the total hydrogen-ionizing emissivity. The redshift \( z_{\rm QSO} \) is the redshift at which QSOs become active, and \( f_{\rm QSO\,haloes} \) is the fraction of halos that host QSOs. The time scale \( \tau_{\rm QSO} \) represents the assumed QSO lifetime. For galaxy-only models, the ionizing spectrum is modelled as a blackbody with temperature \( T_{\rm bb} \), while QSO sources (if present) are assumed to have a power-law spectrum.}
\label{tab:simulations}
\begin{tabular}{lccccccc}
\hline
\textbf{Simulation} &
$\boldsymbol{M_{\rm min}} \, (\mathrm{M}_\odot/h)$ &
\textbf{QSO contribution} &
$\dot{N}_{\rm QSO}/\dot{N}_{\rm total}$ &
$\boldsymbol{z_{\rm QSO}}$ &
$f_{\rm QSO\,haloes}$ &
$T_{\rm QSO}\,(\mathrm{Myr})$ &
$T_{\rm bb}$ \\
\hline
Fiducial         
& $10^9$          
& No             
& --             
& --             
& --             
& --             
& $4\times10^4\,\mathrm{K}$\\
Early  
& $10^9$ 
& No             
& --             
& --             
& --             
& --             
& $4\times10^4\,\mathrm{K}$ \\
Extremely early  
& $10^9$ 
& No             
& --             
& --             
& --             
& --             
& $4\times10^4\,\mathrm{K}$ \\
Oligarchic       
& $8.5\times10^9$
& No             
& --             
& --             
& --             
& --             
& $4\times10^4\,\mathrm{K}$ \\
Democratic       
& $10^9$
& No             
& --             
& --             
& --             
& --             
& $4\times10^4\,\mathrm{K}$ \\
QSO-only         
& $10^9$         
& Yes            
& 100\%          
& 17             
& 10\%           
& 10             
& -- \\
QSO-assisted     
& $10^9$         
& Yes            
& 17\%           
& 10             
& 1\%            
& 10             
& -- \\
MP19
& --        
& --            
& --          
& --             
& --            
& --            
& -- \\
\hline
\end{tabular}
\end{table*}




\begin{figure*}
	\includegraphics[width=0.8\textwidth, trim={0cm 0cm 0cm 0cm}, clip]{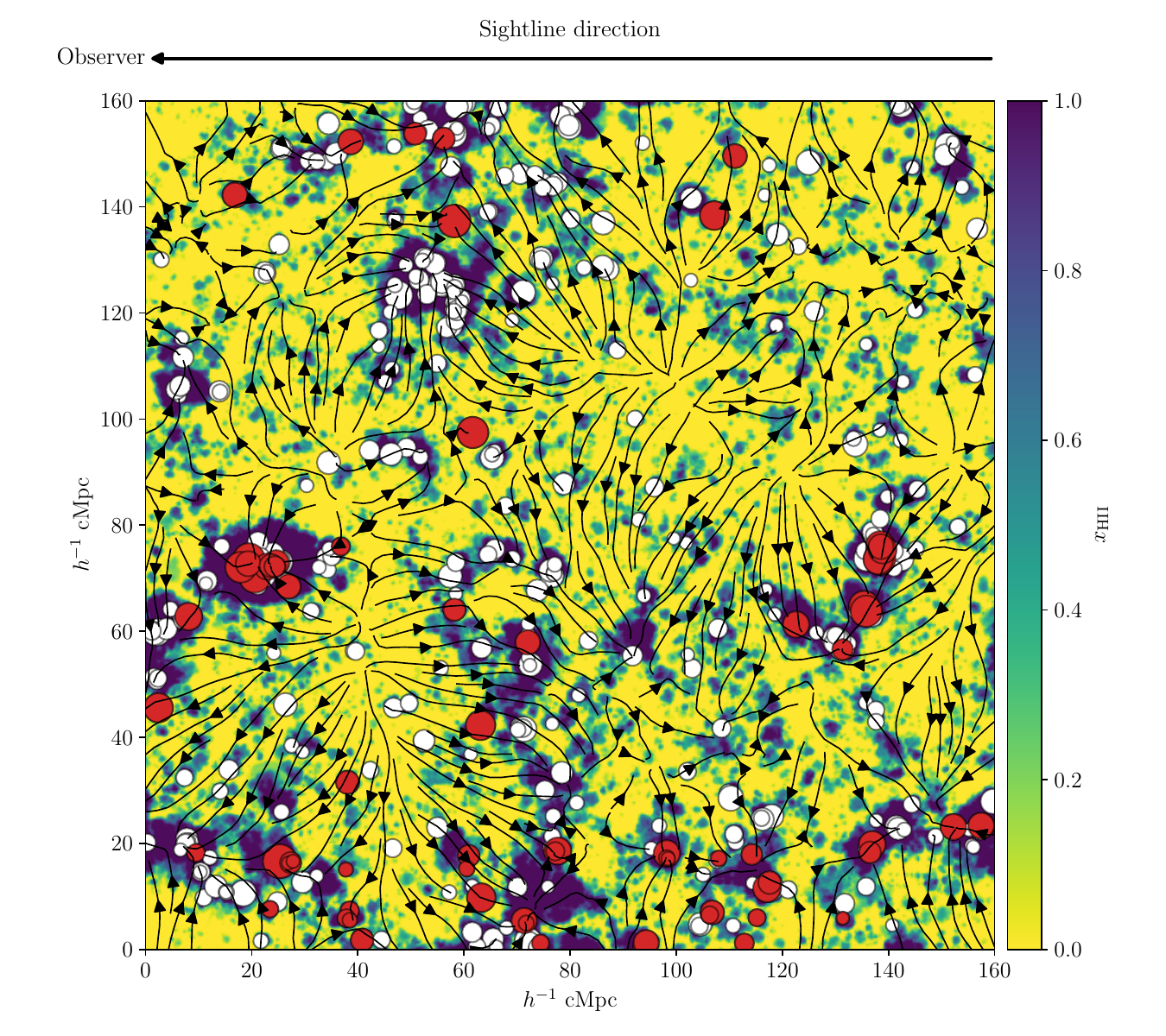}%
	\caption{Ionization structure and galaxy distribution in the Fiducial simulation at $z = 7.14$. The plot shows a slice through the simulation volume depicting the spatial distribution of the ionized hydrogen fraction ($x_{\rm HII}$). The slice is oriented such that the horizontal axis corresponds to the line-of-sight direction from LAEs, with the observer sitting at the left.  The thickness of the slice is $2.5\,h^{-1}$~cMpc, and the transverse dimensions are $(160\,h^{-1} \, \text{cMpc})^2$. White circles denote the locations of Lyman-break galaxies (LBGs) with intrinsic \lya\ luminosities exceeding $10^{42} \, \text{erg} \, \text{s}^{-1}$. Red circles mark the subset of these galaxies identified as LAEs after applying attenuation due to the circumgalactic and intergalactic medium, consistent with the observational selection threshold of $L_{\rm Ly\alpha} > 10^{42} \, \text{erg} \, \text{s}^{-1}$. The black arrows trace the peculiar velocity field within the slice, highlighting the velocity streamlines that generate redshift-space distortions along the line of sight. We see that the peculiar velocity also affects the visitbility of LBGs as LAEs. }

\label{Fig:Slice_plots}
\end{figure*}

\section{A Model for the LAE}
\label{Sec:Methodology} 

To study the clustering of high-redshift LAEs, we populate our simulation boxes with LAEs using a model closely following that introduced by \citet{weinberger2019}. This methodology combines a physically motivated assignment of Lyman-break galaxies (LBGs) to dark matter haloes via abundance matching to the observed UV luminosity function, an empirical model of the intrinsic \lya\ equivalent width distribution, and a computation of \lya\ transmission through the CGM and IGM.
The implementation of this methodology is provided through \textsc{SiMPLE-Gen}\footnote{\textsc{SiMPLE-Gen}: Simulated Mock Population of Lyman-Alpha Emitters Generator. Code available at \url{https://github.com/soumak-maitra/SiMPLE-Gen}.}.
We now describe each component of this modeling framework in detail.

\subsection{Populating Dark matter Haloes with LBGs}
In the first stage of our modeling, we populate the dark matter haloes in our simulation volume with LBG-type galaxies. We adopt a conditional luminosity function method similar to that of \citet{trenti2010}. In this approach, each dark matter halo with mass $M_{\rm h}$ is assumed to host one galaxy. However, the probability that a galaxy is observable at a given time is governed by a duty cycle, $\epsilon_{\rm DC}(M_{\rm h},z)$, which is both mass- and redshift-dependent. This factor accounts for the bursty nature of star formation at high redshift and the fact that galaxies are not continuously luminous in the UV.
The abundance matching is done by requiring that the number density of haloes above a given mass threshold (weighted by the duty cycle) equals the number density of LBGs brighter than a corresponding UV luminosity threshold,
\begin{equation}
  \epsilon_{\rm DC}(M_{\rm h},z)\,\int_{M_{\rm h}}^\infty n(M,z)\,dM \;=\; \int_{L_{\rm UV}}^\infty \phi(L,z)\,dL,
  \label{eq:abundance_match}
\end{equation}
where $n(M,z)$ is the halo mass function. In this work, we adopt the Sheth-Mo-Tormen halo mass function \citep{sheth2001} for $n(M,z)$, and $\phi(L,z)$ is represented by a Schechter function calibrated to the observed UV luminosity functions in \citet{Bouwens2015}, who fitted the Schechter parameters as linear functions of redshift over \(z\sim4\)–8. Equation~(\ref{eq:abundance_match}) thereby defines the mapping $L_{\rm UV}=L_{\rm UV}(M_{\rm h})$.  We write the duty cycle as \citep{trenti2010} 
\begin{equation}
  \epsilon_{\rm DC}(M_{\rm h},z) \;=\; \frac{\displaystyle \int_{M_{\rm h}}^\infty \left[n(M,z)-n(M,z_\Delta)\right]\,dM}{\displaystyle \int_{M_{\rm h}}^\infty n(M,z)\,dM},
  \label{eq:duty_cycle}
\end{equation}
where the time interval $\Delta t$ is given by
\begin{equation}
  \Delta t \;=\; t_{\mathrm H}(z)-t_{\mathrm H}(z_\Delta).
  \label{eq:time_interval}
\end{equation}
Here, \( t_H(z) \) denotes the \textit{Hubble time} at redshift \( z \), defined as the inverse of the Hubble parameter,
\begin{equation}
t_{\mathrm H}(z) = \frac{1}{{\mathrm H}(z)},
\end{equation}
which gives a characteristic timescale for the expansion of the Universe at that redshift. The term \( t_{\mathrm H}(z_\Delta) \) corresponds to the \textit{lookback time} to redshift \( z_\Delta \), i.e., the time elapsed between redshift \( z_\Delta \) and the present day. In this context, \( z_\Delta \) is defined such that the interval \( \Delta t = t_{\mathrm H}(z) - t_{\mathrm H}(z_\Delta) \) represents the recent past (relative to redshift \( z \)) during which significant halo mass growth and star formation are assumed to occur. This interval sets the duty cycle window and quantifies how recently a halo must have formed or assembled mass to be considered actively star-forming. We set $\Delta t = 50\,{\rm Myr}$, a choice motivated by numerical studies that suggest rapid variations in the star formation rate and ionizing photon escape fraction on timescales of 10--100$\,{\rm Myr}$ \citep{rosdahl2018}. This relatively short duty cycle results in a mapping $L_{\rm UV}(M_{\rm h})$ in which lower-mass haloes may contribute significantly to the bright end of the UV luminosity function, and has important implications for the spatial clustering of galaxies.

To accurately mimic observational conditions, the positions of these LBGs are then modified to account for redshift-space distortions (RSD). These distortions arise from the line-of-sight peculiar velocities, $v_{\parallel}$, of the halos, which shift their observed positions according to
\begin{equation}
  \mathbf{s} = \mathbf{x} + \frac{v_{\parallel}}{a{\mathrm H}}\hat{\mathbf{n}},
\end{equation}
where $\mathbf{s}$ is the redshift-space position, $a$ is the scale factor, ${\mathrm H}$ is the Hubble parameter, and $\hat{\mathbf{n}}$ is the line-of-sight direction. 

\subsection{Assigning Intrinsic Ly\texorpdfstring{$\bm{\alpha}$}{alpha} Equivalent Widths}\label{sec:lyalpha_intrinsic}

To select the subset of LBGs based on their Ly$\alpha$ emission, we assign a Ly$\alpha$ rest-frame Equivalent Width (REW) to each LBG, drawn from an empirically motivated probability distribution. We assume that the conditional probability for an LBG with UV magnitude $M_{\rm UV}$ to have an equivalent width REW follows the $M_{\rm UV}$-dependent model proposed by \citet{dijkstra2012}, given by:
\begin{equation}
  P(\mathrm{REW}\,|\,M_{\mathrm{UV}}) \;=\;  \mathcal{N}\,\exp\left[-\frac{\mathrm{REW}}{\mathrm{REW}_\mathrm{c}(M_{\mathrm{UV}})}\right],
  \label{eq:rew_distribution}
\end{equation}
where the normalization $\mathcal{N}$ ensures that the probability integrates to unity over the range $\mathrm{REW}_{\min}(M_{\mathrm{UV}}) \le \mathrm{REW} \le \mathrm{REW}_{\max}$. In our implementation, we adopt $\mathrm{REW}_{\max} = 300\,\text{\AA}$, while the minimum allowed value depends on $M_{\mathrm{UV}}$ as: 
\begin{equation}
  \mathrm{REW}_{\min}(M_{\mathrm{UV}}) =
  \begin{cases}
    -20\,\text{\AA}, & M_{\mathrm{UV}} < -21.5, \\[1mm]
    17.5\,\text{\AA}, & M_{\mathrm{UV}} > -19.0, \\[1mm]
    -20 + 6(M_{\mathrm{UV}}+21.5)^2\,\text{\AA}, & \text{otherwise}.
  \end{cases}
  \label{eq:rew_min}
\end{equation}

Negative values of ${\rm REW}_{\rm min}$ reflect galaxies exhibiting Ly$\alpha$ primarily in absorption rather than emission. These arise from significant attenuation of the UV continuum due to dust or interstellar neutral hydrogen, which can produce absorption features \citep{dijkstra2012}.

The characteristic equivalent width ${\rm REW}_\mathrm{c}$ is assumed to vary with $M_{\rm UV}$ and redshift $z$ as
\begin{equation}
  {\rm REW}_\mathrm{c}(M_{\rm UV}) \;=\; 23 \,+\, 7\,\left(M_{\rm UV}+21.9\right) \,+\, 6\,(z-4).
  \label{eq:rew_char}
\end{equation}
The normalization constant is then given by:
\begin{equation}
    \mathcal{N} = \frac{1}{\mathrm{REW}_c} \left( \exp \left( \frac{-\mathrm{REW}_{\min}}{\mathrm{REW}_c} \right) - \exp \left( \frac{-\mathrm{REW}_{\max}}{\mathrm{REW}_c} \right) \right)^{-1}.
\end{equation}

\citet{dijkstra2012} showed that this prescription provides a good match to observed REW distributions at $z \sim 3.1$, 3.7, and 5.7 \citep{ouchi2008}, although it tends to overpredict high-REW systems at higher redshifts. However, those comparisons did not account for CGM/IGM attenuation. When this attenuation is included, the overprediction is largely alleviated. For instance, \citet{weinberger2019} demonstrated in their Figure~5 that simulated REW distributions using this model, with IGM transmission applied, are in broad agreement with observations. This model also reproduces the REW distribution seen in the Silverrush sample of Ly$\alpha$ emitters.

Nonetheless, caution is required when applying this model to other galaxy populations. For example, recent JWST observations by \citet{tang2023} at $z \sim 5$–6 reveal a significantly different REW distribution compared to that predicted by the Dijkstra \& Wyithe model, suggesting a potentially distinct intrinsic Ly$\alpha$ emission profile. This highlights the importance of using the appropriate unabsorbed EW distribution specific to the galaxy population or survey being modeled, especially when inferring IGM neutral fractions from observed Ly$\alpha$ visibility.

Once a REW is drawn from Equation~\eqref{eq:rew_distribution}, the intrinsic Ly$\alpha$ luminosity is computed as:
\begin{equation}
  L_{\rm Ly\alpha}^{\rm int} \;=\; \frac{\nu_{\alpha}}{\lambda_{\alpha}}\,\left(\frac{\lambda_{\rm UV}}{\lambda_{\alpha}}\right)^{-\beta-2}\times {\rm REW}\times L_{\rm UV,\nu},
  \label{eq:intrinsic_lya}
\end{equation}
where $\nu_{\alpha}=2.47\times10^{15}\,{\rm Hz}$ and $\lambda_{\alpha}=1215.67\,\text{\AA}$ are the frequency and wavelength of the Ly$\alpha$ transition, $\lambda_{\rm UV}$ (typically $1500$--$1600\,\text{\AA}$) is the reference UV wavelength, $\beta\simeq-1.7$ is the assumed UV spectral slope, and $L_{\rm UV,\nu}$ is the monochromatic UV luminosity related to $M_{\rm UV}$ by:
\begin{equation}
  M_{\rm UV} \;=\; -2.5\,\log L_{\rm UV,\nu} + 51.6.
  \label{eq:uv_mag_relation}
\end{equation}

After computing $L_{\rm Ly\alpha}^{\rm int}$, we apply a model for CGM/IGM attenuation and impose observational selection thresholds on both REW and Ly$\alpha$ luminosity to define our LAE sample. The thresholds are based on those adopted in narrow-band surveys, as summarized in Table~\ref{Tab:LAE_selection_threshold}.

\begin{table}
    \centering
    \caption{Observational selection thresholds used for the \lya\ luminosity functions in Fig.~\ref{Fig:LLya}.}\label{Tab:LAE_selection_threshold}
    \begin{tabular}{l c c c}
        \toprule
        \textbf{Survey} & \boldmath{$z$} & \boldmath{REW$_\text{min}$ (\text{\AA})} & \boldmath{$L_{\text{Ly}\alpha,\text{min}}$ (erg s$^{-1}$)} \\
        \midrule
        \citet{Konno2018} & 5.7 & 10 & $6.3 \times 10^{42}$ \\
        \citet{Konno2018} & 6.6 & 14 & $7.9 \times 10^{42}$ \\
        \citet{Ota2017},  & 7.0 & 10 & $2 \times 10^{42}$ \\
     \citet{itoh2018} &  &  &  \\
        \citet{konno2014} & 7.3 & 0 & $2.4 \times 10^{42}$ \\
        \bottomrule
    \end{tabular}
    
    \vspace{2mm}
    \raggedright

\end{table}

\subsection{Ly\texorpdfstring{$\bm{\alpha}$}{alpha} Transmission Through CGM and IGM}
\label{sec:transmission}
The intrinsic Ly$\alpha$ emission emerging from a galaxy is subject to resonant scattering by neutral hydrogen in both the circumgalactic medium (CGM) and the intergalactic medium (IGM). To compute the fraction of Ly$\alpha$ photons transmitted for each LAE in our simulation volume, we extract sightlines through the simulation volume centered on each of the LAE host haloes. Along each sightline, the optical depth $\tau_{\rm Ly\alpha}(v)$ is computed as a function of the velocity offset $v$ from the systemic (line center). We apply the periodic boundary condition of the simulation to shift each sightline, ensuring that the LAE host halo is centered. This allows us to obtain Ly$\alpha$ transmission profiles even for halos located near the edges of the simulation box.

We assume that the intrinsic emission profile $J(v)$ of Ly$\alpha$ photons escaping the galaxy can be described by a Gaussian profile:
\begin{equation}
  J(v) \;=\; \frac{1}{\sqrt{2\pi}\,\sigma_v}\,\exp\!\left[-\frac{\left(v-\Delta v\right)^2}{2\,\sigma_v^2}\right],
  \label{eq:gaussian_profile}
\end{equation}
with a dispersion $\sigma_v=88\,\mathrm{km\,s^{-1}}$ \citep{choudhury2015}. We also introduce a velocity offset $\Delta v$ to account for radiative transfer effects in the galaxy’s interstellar medium (ISM), modeled as:
\begin{equation}
  \Delta v \;=\; a\,v_{\rm circ},
  \label{eq:velocity_offset}
\end{equation}
where $v_{\rm circ}$ is the circular velocity of the host halo and $a = 1.5$ is a proportionality constant, calibrated to match the observed Ly$\alpha$ luminosity function from \citet{Konno2018} at $z=5.756$ (see also Fig.~13 of \citealt{weinberger2019}). This prescription is motivated by observational and theoretical studies showing correlations between halo kinematics and Ly$\alpha$ line profiles \citep[e.g.,][]{neufeld1990, dijkstra2006, verhamme2018}.

While this Gaussian model captures the bulk velocity shift and line width, \citet{yang2017} and \citet{yuan2024} has explored more realistic emergent Ly$\alpha$ line profiles featuring redshifted 'tails' and the role of dust in shaping the equivalent width (EW) distribution. 

Such extensions provide valuable insights into ISM radiative transfer and intrinsic emission complexities. In this work, we adopt a Gaussian profile as a simplified and commonly used approximation to describe the intrinsic Ly$\alpha$ emission, focusing instead on modeling the transmission through the CGM and IGM where large-scale neutral hydrogen dominates the attenuation. {  Our primary aim here is to compare the imprints of the morphology of ionized regions on the bispectrum versus the power spectrum. Using a fixed intrinsic line profile across models ensures that differences in higher-order clustering statistics are driven purely by ionization topology. For future work involving detailed comparisons with observational data, more realistic modeling of the intrinsic line profile will be essential.}

The net transmission fraction, $T_{\rm IGM}$, is then computed as
\begin{equation}
  T_{\rm IGM} \;=\; \frac{\displaystyle \int J(v)\,\exp\!\left[-\tau_{\rm Ly\alpha}(v)\right]\,dv}{\displaystyle \int J(v)\,dv}.
  \label{eq:transmission_fraction}
\end{equation}
Here, $\tau_{\rm Ly\alpha}(v)$ is the optical depth along the sightline bluewards of the LAE host halo redshift. 
The observed Ly$\alpha$ luminosity is then determined by applying the transmission fraction to the intrinsic luminosity:
\begin{equation}
  L_{\rm Ly\alpha}^{\rm obs} \;=\; T_{\rm IGM}\,L_{\rm Ly\alpha}^{\rm int}.
  \label{eq:observed_lya}
\end{equation}
This gives us the \lya\ emitter luminosity for the galaxies that one observes accounting for the CGM/IGM attenuation.

In Fig.~\ref{Fig:Slice_plots}, we present a 2.5~$h^{-1}$cMpc-thick slice through the simulation volume, oriented such that the sightline to the observer is along the horizontal axis, with the observer positioned at redshift $z = 0$. The colour scale depicts the number density of neutral hydrogen, $n_{\rm HI}$, averaged over 2.5~$h^{-1}$cMpc along the depth of the slice. This slice corresponds to redshift $z = 7.14$ and assumes the Fiducial reionization history. LAEs with intrinsic Ly$\alpha$ luminosities \( L_{\rm Ly\alpha}^{\rm int} > 10^{42} \) erg s$^{-1}$ are marked with white circles, while those that remain above this threshold in observed luminosity after attenuation by the circumgalactic (CGM) and intergalactic medium (IGM) are shown in red. The area of each circle is proportional to its corresponding Ly$\alpha$ luminosity. To illustrate gas dynamics, we overlay streamlines of the peculiar velocity field within the slice, emphasizing local gas flows capable of inducing redshift-space distortions along the line of sight and thereby affecting the detectability of LAEs.

In general, LAEs are located within extended ionized regions, but their detectability is governed by a complex interplay between intrinsic luminosity, the ionization structure of the surrounding medium, and local velocity gradients. For instance, near coordinates $(50,125)\ h^{-1}$cMpc, several LAEs reside within an apparently ionized bubble, yet none satisfy our detection criteria. This can be attributed to strong infall velocities that compress the ionization front and suppress the emerging Ly$\alpha$ flux through increased damping of the line profile. These conditions prevent the Ly$\alpha$ photons from escaping efficiently, keeping the observed luminosity below threshold. Thus, the probability of detecting an LAE is not merely a function of its intrinsic output, but also of its placement within the patchy reionization topology and the velocity-induced distortions of its Ly$\alpha$ emission.

\begin{figure*}
	\includegraphics[width=6.55cm, trim={0.5cm 0cm 0.67cm 0.2cm}, clip]{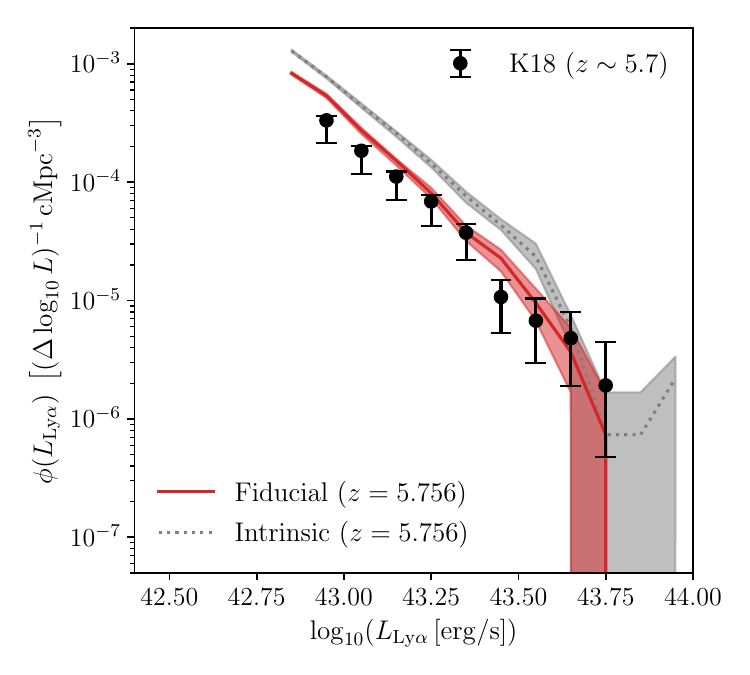}%
    \includegraphics[width=5.57cm,trim={2.2cm 0cm 0.67cm 0.2cm}, clip]{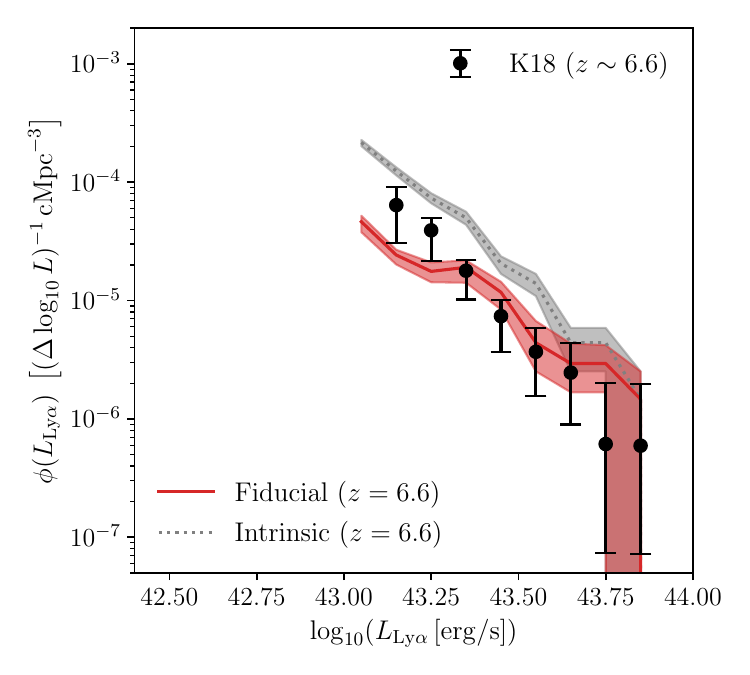}%
    \includegraphics[width=5.57cm,trim={2.2cm 0cm 0.67cm 0.2cm}, clip]{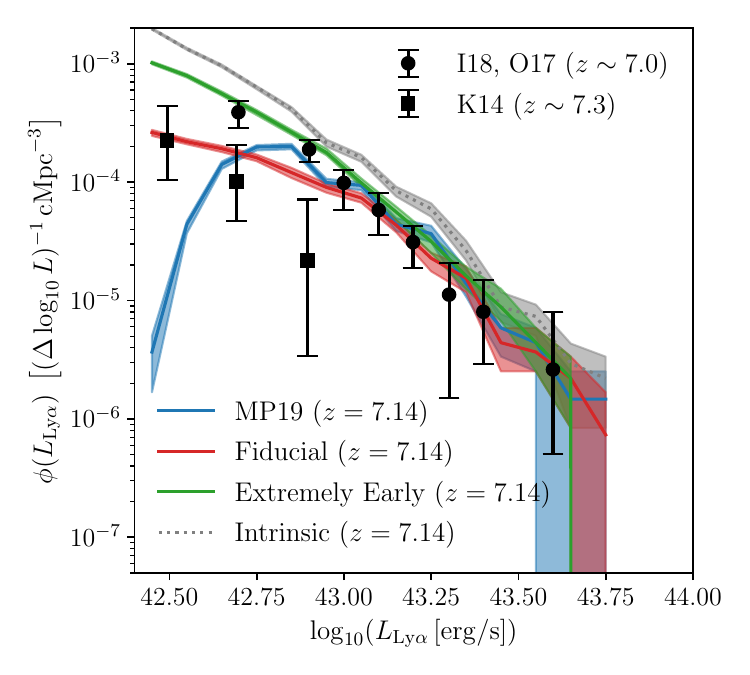}%
	\caption{\lya\ Luminosity function of simulated mock LAEs for $z=5.756, 6.6$  and 7.14. The red curve plots our fiducial model for the observed \lya\ luminosity accounting for attenuation due to the CGM/IGM. We compare it with the observed \lya\ luminosity function of LAEs plotted as black markers with error bars, observed using narrow-band imaging data acquired through the NB816, NB921, NB973, and NB101 filters (corresponding to redshift of the observable Ly$\alpha$ emission lines at $z_{\rm Ly\alpha}= 5.72\pm 0.046, 6.58\pm 0.056, 6.99\pm 0.046$ and $7.30\pm 0.037$) on the Hyper Suprime-Cam (HSC) instrument mounted on the Subaru Telescope \citep{konno2014,Konno2018,Ota2017,itoh2018}. We also plot the intrinsic \lya\ Luminosity function without accounting for the CGM/IGM attenuation with a dotted grey curve. }
\label{Fig:LLya}
\end{figure*}
 
\subsection{LAE Luminosity Function}

We compare our simulated \lya\ emitter luminosity functions at redshifts $z=5.756$, $6.6$, and $7.14$ with observational data in Figure~\ref{Fig:LLya}. The simulations are based on our original $(160 h^{-1}\text{cMpc})^3$ volume, with error bars estimated via jackknife resampling from eight sub-volumes. These error bars represent 68\% confidence intervals derived from the jackknife method.

For observational comparisons, we adopt measurements from narrow-band imaging surveys carried out with the NB816, NB921, NB973, and NB101 filters on the Hyper Suprime-Cam (HSC) instrument mounted on the Subaru Telescope \citep{konno2014, Konno2018, Ota2017, itoh2018}. These filters are sensitive to Ly$\alpha$ emission lines at redshifts of $z_{\rm Ly\alpha} = 5.72 \pm 0.046$, $6.58 \pm 0.056$, $6.99 \pm 0.046$, and $7.30 \pm 0.037$, respectively. Since our simulation snapshots are saved at 40~Myr intervals and do not precisely align with these observational redshifts, we make the following comparisons: the simulated luminosity function at \( z = 7.14 \) is compared with observations at \( z_{\rm Ly\alpha} = 6.99 \pm 0.046 \) and \( z_{\rm Ly\alpha} = 7.30 \pm 0.037 \); the simulated luminosity function at \( z = 6.6 \) is compared with observations at \( z_{\rm Ly\alpha} = 6.58 \pm 0.056 \); and the simulated luminosity function at \( z = 5.756 \) is compared with observations at \( z_{\rm Ly\alpha} = 5.72 \pm 0.046 \). The corresponding selection thresholds used for our mock LAE samples are listed in Table~\ref{Tab:LAE_selection_threshold}.

For the $z=5.756$ mock LAE sample, we apply a minimum selection threshold of $L_{\text{Ly}\alpha,\text{min}} > 6.3 \times 10^{42}$ erg s$^{-1}$  and a REW cut of REW$_\text{min} > 10$ \AA, based on the criteria from \citet{Konno2018} at $z = 5.72 \pm 0.046$. The resulting luminosity function from our Fiducial model slightly overpredicts the LAE number density at low luminosities but is broadly consistent with observations within the quoted uncertainties.

For the $z=6.6$ mock sample, we adopt selection thresholds of $L_{\text{Ly}\alpha,\text{min}} > 7.9 \times 10^{42}$ erg s$^{-1}$ and REW$_\text{min} > 14$ \AA, consistent with observational data at $z = 6.58 \pm 0.056$ \citep{Konno2018}. At this redshift, our Fiducial model reproduces the observed luminosity function remarkably well within the observational uncertainties.

For the $z=7.14$ sample, we impose selection thresholds of $L_{\text{Ly}\alpha,\text{min}} > 2 \times 10^{42}$ erg s$^{-1}$ and REW$\text{min} > 4$ \AA, in line with the observational criteria reported at $z = 6.99 \pm 0.046$ \citep{Ota2017, itoh2018}. We also compare our results with observations at a slightly higher redshift, $z = 7.30 \pm 0.037$ \citep{konno2014}. Across all models considered—Fiducial, MP19, and Extremely Early reionization—the simulations agree well with observations at high luminosities ($L_{\text{Ly}\alpha} > 10^{43}$ erg s$^{-1}$). At lower luminosities, the predictions exhibit significant scatter. The Extremely Early reionization model matches the $z = 6.99 \pm 0.046$ observations most closely, while the Fiducial model better aligns with the $z = 7.30 \pm 0.037$ data.
The MP19 model, on the other hand, displays a sudden decline at the faint end of the luminosity function. This is likely due to the homogeneous ionizing background in MP19, which overexposes faint galaxies to Ly$\alpha$ attenuation, suppressing their visibility compared to models with patchy reionization. As shown in the right panel of Fig.~\ref{Fig:LLya}, this results in an underprediction of faint LAEs. This differential behavior between the bright and faint LAEs (minimal evolution at the bright end but strong, model-dependent suppression at the faint end) emphasizes the power of the luminosity function as a probe of the patchiness and timing of reionization. By tracking how the faint population is differentially quenched relative to the bright population, one can place tighter constraints on the spatial distribution of neutral hydrogen and the progression of ionized regions during the Epoch of Reionization \citep{Weinberger2018}.

Overall, the close agreement between our simulated luminosity functions and the HSC observational data underscores the reasonableness of the LAE modelling described in this section.

\section{Three-dimensional power spectrum and bispectrum of LAEs}
\label{Sec:Clustering}

We now quantify the 3D clustering properties of LAEs using the 3D power spectrum $P(k)$, and bispectrum $B(k_1,k_2,k_3)$, which describe the two-point and three-point statistics, respectively. While the power spectrum quantifies the variance of a fluctuating field as a function of scale, the bispectrum is sensitive to its non-Gaussian features. It is well-known that the spatial clustering of LAEs serves as a powerful probe of both the underlying cosmological density field and the ionization field \citep{Greig2013,Hutter2015,weinberger2019}. After presenting the power spectrum and the bispectrum of LAEs in our reionization models, we will compare the performance of the two statistics in probing the characteristic scales and morphology of the ionization field at a given epoch.

\subsection{LAE number density} 

LAEs are discrete tracers of the underlying density field and ionization field. To compute the 3D power spectrum and bispectrum, one first needs to map these discrete tracers to a 3D field. The construction of this 3D field begins with the placement of our modeled LAEs onto a uniform grid. For our simulation volume of side-length $L = 160\,h^{-1}\mathrm{cMpc}$, we subdivide it into $512^3$ grids, which sufficiently resolves the smallest scales that we probe with power spectrum and bispectrum in this work. The LAE positions are assigned to these grids using the Nearest-Grid-Point (NGP) mass-assignment scheme. 
Following this mass assignment, we obtain the resulting number count of LAEs at each grid $n(\mathbf{x})$. We then normalise the numbers $n(\mathbf{x})$ by the mean number density, $\bar{n}$, to obtain the overdensity field of the discrete tracers, which in this case are the visible LAEs,
\begin{equation}
  \delta(\mathbf{x}) = \frac{n(\mathbf{x})}{\bar{n}} - 1.
\end{equation}
For computing the power spectrum and bispectrum along with the associated uncertainties, we first divide our simulation volume $(160\,h^{-1}\mathrm{cMpc})^3$ into 27 equal sub-volumes of dimensions $(80\,h^{-1}\mathrm{cMpc})^3$ each distributed uniformly in the box with possible spatial overlap. We then compute the power spectrum and bispectrum individually for these sub-volumes and estimate their mean values. The sub-volumes allow us to assign uncertainties to the estimated power spectrum and bispectrum.

\subsection{Power Spectrum}

The overdensity field is analyzed in Fourier space by applying a Fast Fourier Transform (FFT) as
\begin{equation}
  \delta(\mathbf{k}) = \frac{1}{L^3} \int \delta(\mathbf{x}) \, e^{-i \mathbf{k}\cdot\mathbf{x}} \, \dd^3x.
\end{equation}
After applying the Fourier transform, the 3D power spectrum is expressed as 
\begin{equation}
    (2\pi)^3 P(\mathbf{k},\mathbf{k}') \delta_{\rm D}(\mathbf{k} - \mathbf{k}')
\equiv \langle \delta(\mathbf{k}) \delta(\mathbf{k}') \rangle,
\end{equation}
where $\delta_D$ is the Dirac delta function. 

Numerically, the power spectrum can be computed as
\begin{equation}
    P(k) = \frac{V}{N_k} \sum_{\substack{|\mathbf{k}' - k| < \Delta k}} |\delta(\mathbf{k}')|^2,
\end{equation}

where $N_k$ is the number of Fourier modes in a particular $k$ bin and $V$ is the volume of the simulation box. 

\begin{figure*}
    \includegraphics[width=6.3cm, trim={0.4cm 1.5cm 0.4cm 0.4cm}, clip]{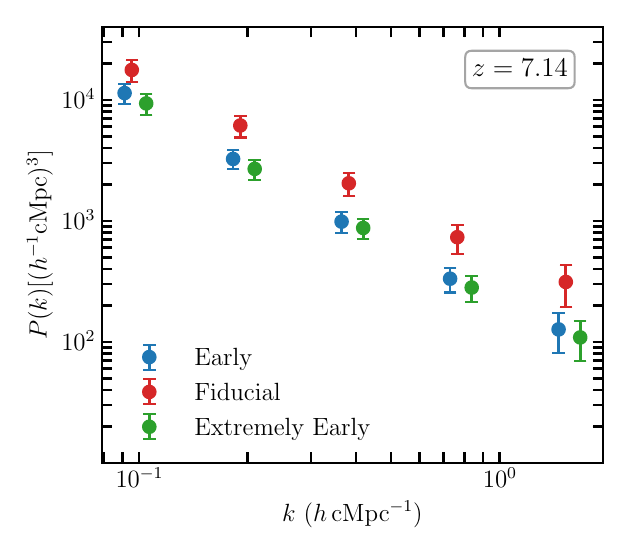}%
    \includegraphics[width=5.5cm, trim={1.65cm 1.5cm 0.4cm 0.4cm}, clip]{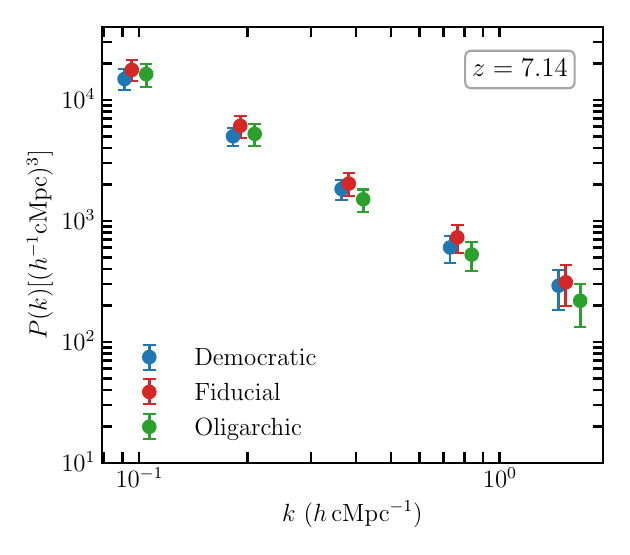}%
    \includegraphics[width=5.5cm, trim={1.65cm 1.5cm 0.4cm 0.4cm}, clip]{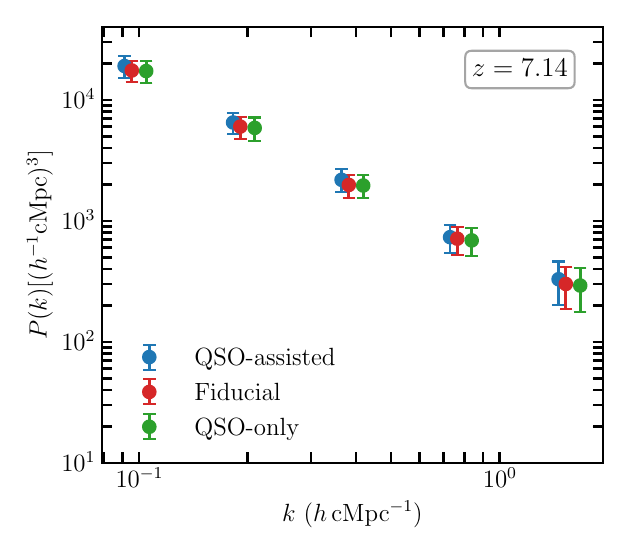}%
    
    \includegraphics[width=6.3cm, trim={0.4cm 0cm 0.4cm 0.4cm}, clip]{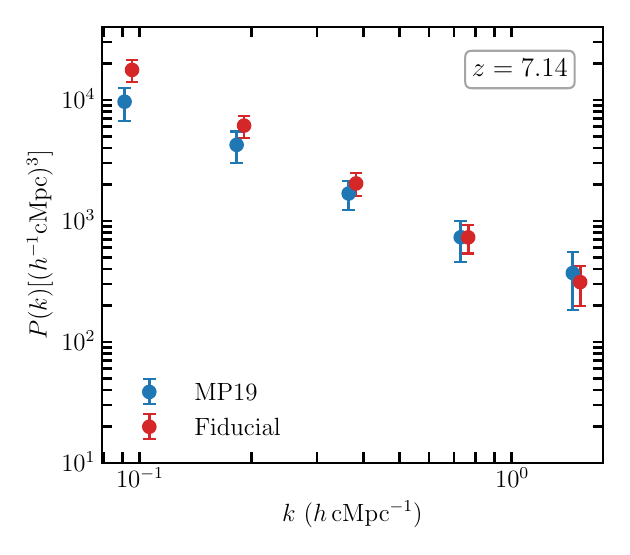}%
    \includegraphics[width=5.5cm, trim={1.65cm 0cm 0.4cm 0.4cm}, clip]{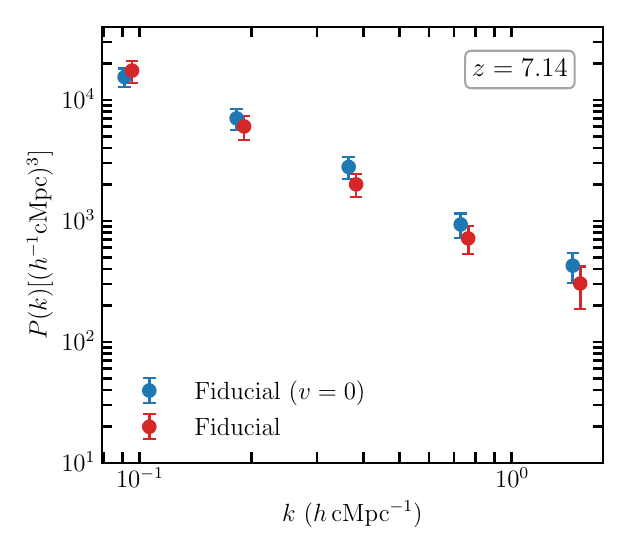}%
    \includegraphics[width=5.5cm, trim={1.65cm 0cm 0.4cm 0.4cm}, clip]{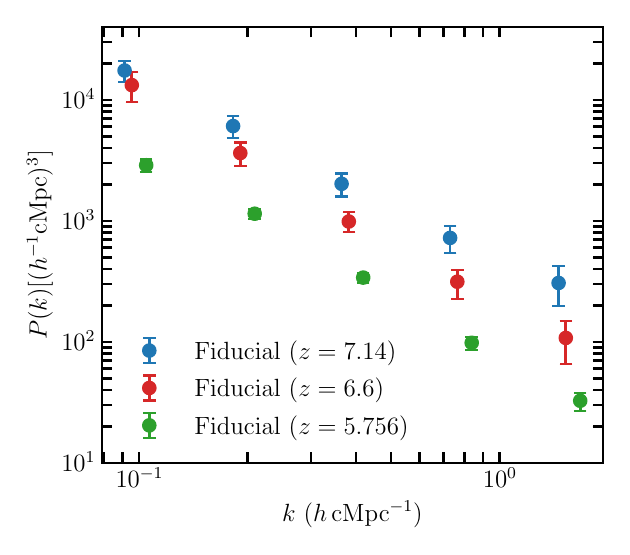}%
    
	\caption{
The 3D power spectrum $P(k)$ of our modeled \lya\ emitters, accounting for CGM/IGM attenuation, at $z=7.14$ unless otherwise noted. 
Each panel compares the Fiducial model with alternative reionization scenarios. 
The top left panel compares the Early, Fiducial, and Extremely Early models; the top middle compares the Democratic, Fiducial, and Oligarchic models; and the top right compares the QSO-assisted, Fiducial, and QSO-only models. 
The bottom left panel compares MP19 and Fiducial, both calibrated to the same $\langle x_{\mathrm{HI}} \rangle$ but differing in the spatial structure of the ionizing background. 
The bottom middle panel compares the Fiducial model with and without redshift-space distortions. 
The bottom right panel shows the redshift evolution of the Fiducial model at $z=5.756$, $6.6$, and $7.14$. 
Error bars represent the $68\%$ confidence intervals estimated via bootstrap resampling of 8 sub-volumes drawn from 27 overlapping regions of size $(80\,h^{-1}\mathrm{cMpc})^3$, such that each bootstrap realization spans the full effective volume of $(160\,h^{-1}\mathrm{cMpc})^3$.
}

\label{Fig:PS}
\end{figure*}

\begin{figure*}
    \includegraphics[width=14.3cm, trim={0.4cm 2.28cm 0cm 7.2cm}, clip]{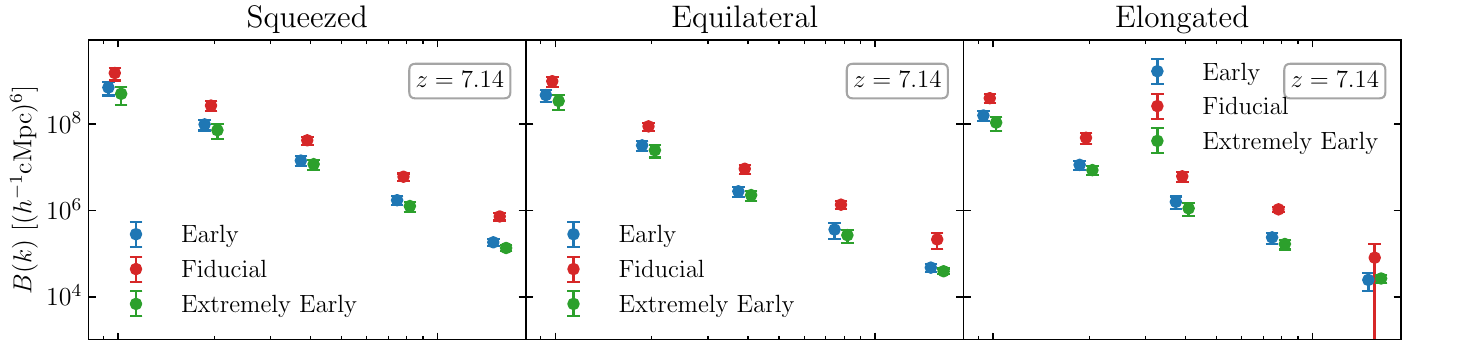}    
    \includegraphics[width=14.3cm, trim={0.4cm 2.28cm 0cm 7.8cm}, clip]{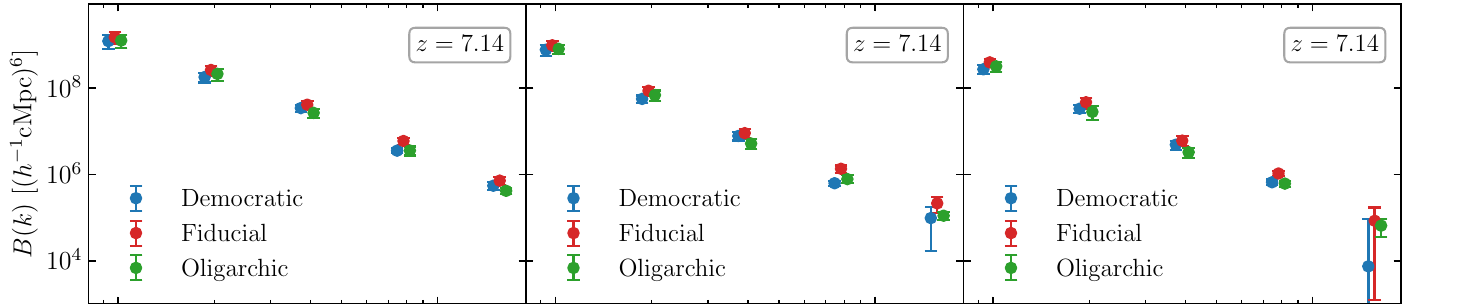}    
    \includegraphics[width=14.3cm, trim={0.4cm 2.28cm 0cm 7.8cm}, clip]{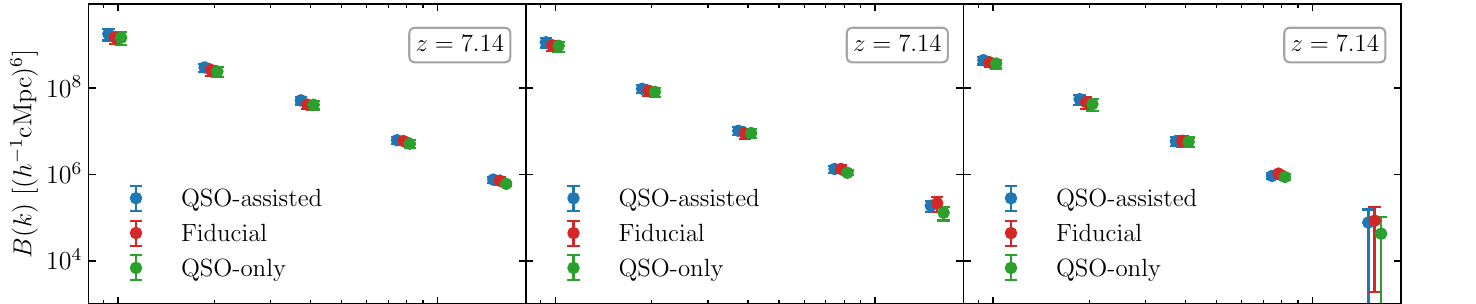}  
    \includegraphics[width=14.3cm, trim={0.4cm 2.28cm 0cm 7.8cm}, clip]{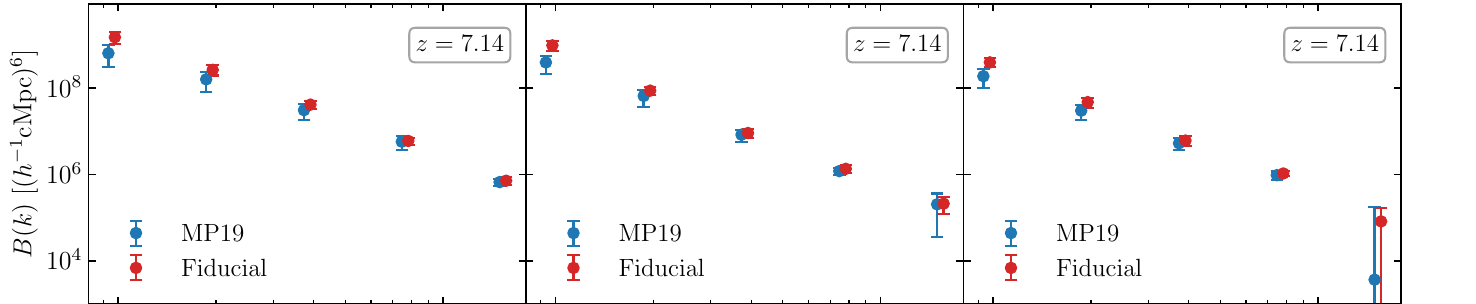}  
    \includegraphics[width=14.3cm, trim={0.4cm 2.28cm 0cm 7.8cm}, clip]{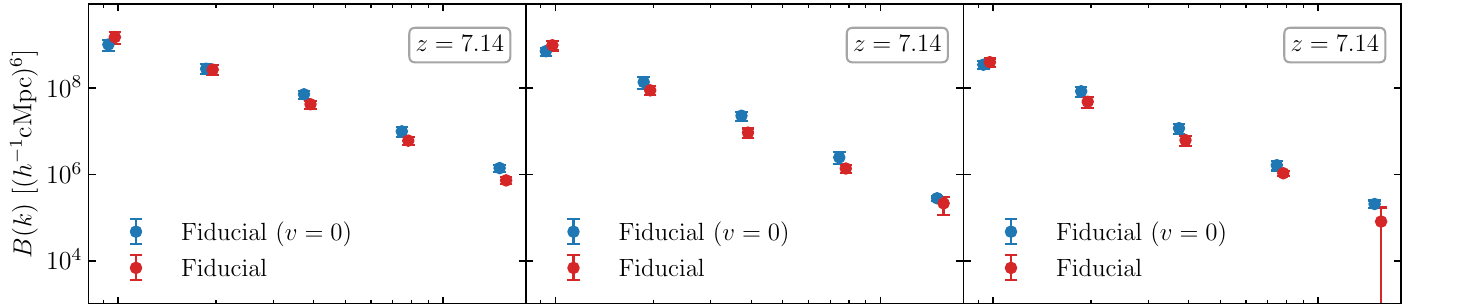}  
    \includegraphics[width=14.3cm, trim={0.4cm 1cm 0cm 7.8cm}, clip]{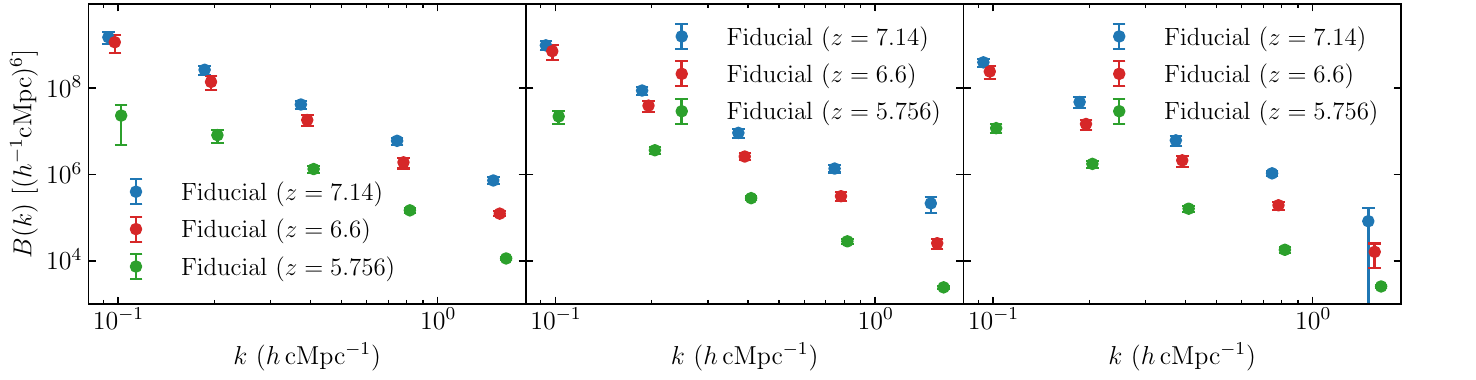}  
	\caption{
The 3D bispectrum $B(k)$ of our modeled \lya\ emitters, accounting for CGM/IGM attenuation, shown for isosceles triangular configurations in $k$-space. 
The left, middle, and right columns correspond to squeezed, equilateral, and elongated triangle shapes, respectively. 
The top row compares the Early, Fiducial, and Extremely Early models at $z=7.14$; 
the second row shows the Democratic, Fiducial, and Oligarchic models; 
the third row presents the QSO-assisted, Fiducial, and QSO-only models. 
The fourth row compares the MP19 and Fiducial models, both calibrated to the same $\langle x_{\mathrm{HI}} \rangle$ but differing in spatial ionization structure. 
The fifth row compares the Fiducial model with and without redshift-space distortions. 
The sixth row displays the redshift evolution of the Fiducial model at $z = 5.756$, $6.6$, and $7.14$. 
Error bars represent the $68\%$ confidence intervals obtained by bootstrap resampling of eight randomly drawn sub-volumes from a total of 27 overlapping sub-volumes of size $(80\,h^{-1}\mathrm{cMpc})^3$, such that each bootstrap sample spans the full simulation volume $(160\,h^{-1}\mathrm{cMpc})^3$. Due to the large dynamic range of $B(k)$, visual differences between models may be subtle, but are quantified more robustly in Section~\ref{Sec:Chi_sqaure} and Fig.~\ref {fig:chisq} through a comparison over all triangular configurations.
}

\label{Fig:BS}
\end{figure*}

Due to the discrete sampling of the density field, there exists an additional contribution from shot noise. The shot noise in the power spectrum is estimated as
\begin{equation}
  P_{\rm shot} = \frac{1}{\bar{n}}.
\end{equation}
(We present a derivation of this shot noise term  in Appendix~\ref{A:Shot_noise}.) The shot noise subtracted power spectrum is then expressed as :
\begin{equation}
  P(k) = \frac{V}{N_k} \sum_{\substack{|\mathbf{k}' - k| < \Delta k}} |\delta(\mathbf{k}')|^2 - P_{\rm shot}.
\end{equation}
We compute the power spectrum, $P(k)$, of the LAE overdensity field using the {\sc Pylians} package \citep{Pylians}. The calculation employs 5 logarithmically spaced bins in wavenumber, $k$, ranging from $0.1$ to $1.6~h\,\text{cMpc}^{-1}$. The power spectrum is derived separately for each of the 27 sub-volumes of our simulation box, as previously described. In addition, we accounted for the Poissonian uncertainty in the number of LAEs by drawing Poisson‐distributed random realizations centered on the measured number of LAEs for each sub‐volume and propagating these through the shot‐noise subtraction.
The mean and corresponding $68\%$ confidence intervals for $P(k)$ are obtained through bootstrap resampling of these sub-volumes.

We perform this procedure consistently across all the reionization models described in Section~\ref{Sec:Simulation}, namely: Fiducial, MP19, Early, Extremely Early, Democratic, Oligarchic, QSO-assisted, and QSO-only. The resulting LAE power spectra for these reionization histories at redshift $z = 7.14$ are shown in the panels of Fig.~\ref{Fig:PS}.

The top left panel compares the Early, Fiducial, and Extremely Early models. This comparison highlights the effect of varying the mean neutral hydrogen fraction ($\langle x_{\rm HI} \rangle$). The Fiducial model, characterized by a higher $\langle x_{\rm HI} \rangle$ than the Early and Extremely Early models, exhibits a noticeably lower clustering amplitude. These models show clear differences in $P(k)$ across a range of scales, suggesting that the power spectrum effectively captures variations in the global ionization state.

The top middle and top right panels explore the impact of different source prescriptions on the power spectrum. The top middle compares the Democratic, Fiducial, and Oligarchic models, while the top right contrasts the QSO-assisted, Fiducial, and QSO-only models. In both cases, the three models appear broadly consistent within their error bars. This suggests that, under our modeling assumptions, the power spectrum exhibits limited sensitivity to the nature of ionizing sources.

The bottom left panel compares the Fiducial and MP19 models. These two models are calibrated to the same mean neutral fraction ($\langle x_{\rm HI} \rangle$) but differ in the spatial structure of the ionizing background: MP19 assumes a spatially uniform UV background, while the Fiducial model includes inhomogeneous, patchy reionization. Although MP19 is not a physically motivated model, it is introduced here as a controlled baseline to assess the impact of ionization patchiness. While differences in $P(k)$ are visible at large scales ($k < 0.2\,h\,\mathrm{cMpc}^{-1}$), the overall separation is modest, and the MP19 model exhibits larger uncertainties, particularly at small scales. This highlights the limitations of the power spectrum in detecting morphological differences when $\langle x_{\rm HI} \rangle$ is fixed.

The bottom middle panel illustrates the effect of redshift-space distortions by comparing the Fiducial model computed with and without peculiar velocities. While the differences are relatively small, redshift-space distortions tend to slightly suppress the amplitude of the power spectrum, especially at intermediate and small scales.

Finally, the bottom right panel shows the redshift evolution of the Fiducial model at $z = 5.756$, $6.6$, and $7.14$. We observe that LAEs exhibit stronger clustering at higher redshifts. This trend is consistent with the picture that, at earlier times, LAEs preferentially reside in isolated, highly clustered ionized regions. As reionization progresses, ionized bubbles grow and merge, leading to a more diffuse ionization topology and reduced LAE clustering amplitude. These trends are in line with earlier findings \citep{Mcquinn2007b,weinberger2019}.

\subsection{Bispectrum}

The 3D bispectrum of the LAE overdensity field $\delta(\mathbf{r})$ is expressed as
\begin{equation}
  (2\pi)^3 \delta_D(\mathbf{k}_1 + \mathbf{k}_2 + \mathbf{k}_3) B(k_1,k_2,k_3)\equiv \langle \delta(\mathbf{k}_1) \delta(\mathbf{k}_2) \delta(\mathbf{k}_3) \rangle.
\end{equation}
Numerically, the bispectrum is computed as 
\begin{equation}
    B(k_1, k_2, k_3) = \frac{V^2}{N_{\rm tri}} \sum_{{\bf k}_1 + {\bf k}_2 + {\bf k}_3 = 0} \tilde\delta({\bf k}_1) \tilde\delta({\bf k}_2) \tilde\delta({\bf k}_3),
\end{equation}
where $N_{\rm tri}$ is the number of triangular Fourier modes in a given bin.
In practice, the bispectrum is often computed for a set of Fourier space triangular configurations parameterized by two side lengths, $k_1$ and $k_2$, and the enclosed angle $\theta$. 

Analogous to the power spectrum, the bispectrum is contaminated by shot noise due to the discreteness of tracers. The shot noise contribution is estimated as
\begin{equation}\label{Eq:B_shotnoise}
  B_{\rm shot}(k_1,k_2,k_3) = \frac{P_{\rm shot}(k_1)}{\bar{n}} + \frac{P_{\rm shot}(k_2)}{\bar{n}} + \frac{P_{\rm shot}(k_3)}{\bar{n}} + \frac{1}{\bar{n}^2}.
\end{equation}
(We derive this expression in Appendix~\ref{A:Shot_noise}.) The measured bispectrum is then corrected by subtracting this contribution.

{ It is important to note that the bispectrum, being a three-point statistic, naturally includes information already encoded in the power spectrum through its dependence on second-order correlations. Therefore, any interpretation of differences in the bispectrum across models implicitly incorporates differences in the underlying two-point clustering as well. Isolating genuinely higher-order (non-Gaussian) information requires caution, especially when the bispectrum signal correlates strongly with the power spectrum amplitude.}

We compute the bispectrum over a grid of triangular configurations parameterized by the lengths of two wavevectors \(k_1\) and \(k_2\), and the angle \(\theta\) between them. The wavenumbers \(k_1\) and \(k_2\) are independently binned into 5 logarithmically spaced intervals spanning the range \(0.1\) to \(1.6\,h\,\mathrm{cMpc}^{-1}\), while \(\theta\) is divided into 9 uniform bins between \(0.13\pi\) and \(0.93\pi\), resulting in 225 triangular configurations (\(5 \times 5 \times 9\)).
To account for sample variance, we compute the bispectrum separately across 27 sub-volumes of our simulation box and obtain the mean and 68\% confidence intervals via bootstrap resampling, with a sample size of 8 to match the volume of the original simulation box. Poisson uncertainty due to discrete LAEs is also included by generating Poisson-distributed realizations of the LAE number in each sub-volume and propagating the resulting variation through the shot-noise subtraction.
We apply this methodology consistently across multiple reionization models: Fiducial, MP19, Early, Extremely Early, Democratic, Oligarchic, QSO-assisted, and QSO-only. The full set of bispectrum values for all 225 triangular configurations is shown in Fig.~\ref{Fig:BS_triangleindex}, and the corresponding configurations are listed in Table~\ref{tab:triplets} in the appendix. Subtracted shot-noise contributions for each triangular index are also plotted.

\subsubsection{Effect of Reionization History and Source Models}

In Fig.~\ref{Fig:BS}, the left, middle, and right columns illustrate the bispectra for squeezed, equilateral, and elongated configurations, respectively. Each is defined by $k_1 = k_2$, with the squeezed shape corresponding to the smallest angle $\theta = 0.13\pi$. The bispectrum amplitudes span values from $10^4$ to $10^7\, (h^{-1}\mathrm{cMpc})^6$, with larger amplitudes at lower $k$, indicative of stronger non-Gaussianity on larger scales.
The top row of Fig.~\ref{Fig:BS} compares the Early, Fiducial, and Extremely Early models. The Fiducial model, having a higher mean neutral hydrogen fraction $\langle x_{\rm HI} \rangle$ than the Early and Extremely Early models, yields a stronger bispectrum signal. This trend is consistent across all configurations, suggesting that the bispectrum, like the power spectrum, effectively captures variations in $\langle x_{\rm HI} \rangle$.
The second and third rows compare the Democratic vs. Oligarchic and QSO-assisted vs. QSO-only models, respectively. Similar to the power spectrum, these models exhibit overlapping bispectra within error bars. However, the bispectrum provides slightly enhanced visual distinction between the Fiducial and Democratic/Oligarchic models, especially in the equilateral and elongated configurations. A more quantitative comparison will be presented in the following subsection.

\subsubsection{Effect of Ionization Morphology: Patchy vs. Uniform UVB}

The fourth row of Fig.~\ref{Fig:BS} contrasts the Fiducial and MP19 models, both calibrated to the same mean neutral hydrogen fraction, $\langle x_{\rm HI} \rangle$. The Fiducial model incorporates a spatially inhomogeneous reionization field, whereas MP19 assumes a spatially uniform UV background. While MP19 is not a physically motivated scenario, it serves here as a diagnostic baseline to isolate the effects of spatial patchiness in the ionizing field.
At first glance, the bispectra of the two models may appear broadly similar across configurations, especially given the large dynamic range of $B(k)$ values. It is therefore not straightforward to assess their distinguishability by eye. However, as we demonstrate quantitatively in Section~\ref{Sec:Chi_sqaure}, a reduced chi-square analysis across all triangle configurations reveals a statistically significant difference between the two. This suggests that, despite the visual similarities, the bispectrum retains sensitivity to differences in ionization morphology that are less evident in the power spectrum.

\subsubsection{Effect of Redshift-Space Distortions}

In the fifth row, we assess the influence of peculiar velocities by comparing the bispectrum of the Fiducial model with and without redshift-space distortions. While velocity effects induce relatively modest changes in the power spectrum, they produce more pronounced suppression in the bispectrum amplitude at intermediate-to-small scales ($k > 0.2\, h\, \mathrm{cMpc}^{-1}$). This is most visible in the squeezed and equilateral configurations. These results underscore the importance of accurately modeling redshift-space effects when interpreting LAE bispectrum measurements.

\subsubsection{Redshift Evolution of the LAE Bispectrum}

The bottom row of Fig.~\ref{Fig:BS} shows the redshift evolution of the bispectrum for the Fiducial model at $z=5.756$, 6.6, and 7.14. Similar to the power spectrum trends, the bispectrum amplitude increases with redshift across all triangle types and scales. This reflects stronger clustering of LAEs at earlier times, consistent with their preferential occupation of rare, highly biased ionized regions during the early stages of reionization. The bispectrum thus provides complementary insights into both the non-Gaussianity and the evolving spatial topology of ionized regions.

\begin{figure*}
\centering
    \includegraphics[width=18cm, trim={0.4cm 0.4cm 0cm 0cm}, clip]{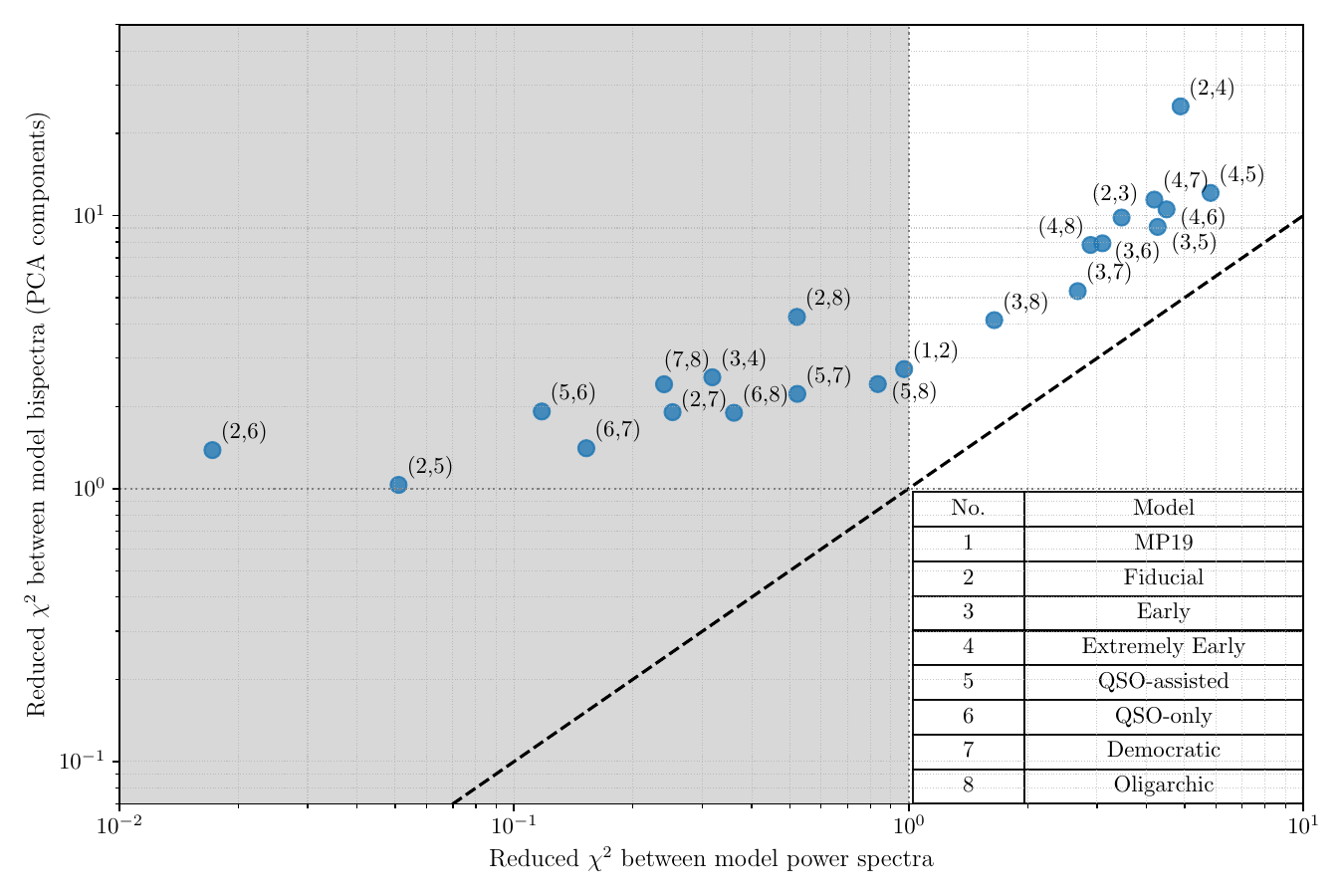}
    \caption{Discriminatory power of power spectrum and bispectrum statistics across reionization model pairs, quantified via reduced $\chi^2$. Each point represents a unique pairwise comparison among the eight reionization models considered in this work, with model indices corresponding to those listed in the accompanying table. Reduced $\chi^2$ values are computed using full covariance matrices derived from 5000 bootstrap realizations. For the bispectrum, given the high dimensionality of the configuration space ($5 \times 5 \times 9$), we perform a Principal Component Analysis (PCA) and retain the top 30 components, which capture the dominant modes of variation, for the $\chi^2$ evaluation. The dashed diagonal line marks equal discriminative power between the power spectrum and bispectrum. Points lying above this line correspond to model pairs that are more effectively distinguished by the bispectrum. The shaded region denotes comparisons where the power spectrum yields reduced $\chi^2 < 1$, indicating limited discriminatory power. The MP19 model is compared only with the Fiducial model to isolate the impact of patchiness, as both share the same mean neutral fraction but differ in the spatial structure of the ionizing background.}

    \label{fig:chisq}
\end{figure*}

\subsection{Quantifying Clustering-Based Distinctions Between Reionization Scenarios}
\label{Sec:Chi_sqaure}

We now carry out a quantitative comparison of the discriminative capabilities of the LAE power spectrum $P(k)$ and bispectrum $B(k)$ in distinguishing various reionization models through a reduced $\chi^2$ analysis.  

We analyze all eight reionization models considered in this work, which are:  
(1) MP19 (uniform UV background),  
(2) Fiducial (Late reionization),  
(3) Early reionization,  
(4) Extremely Early reionization,  
(5) QSO-assisted,  
(6) QSO-only,  
(7) Democratic, and  
(8) Oligarchic models.  
Each model encapsulates distinct ionization morphologies and underlying astrophysical processes.  
Our primary goal is to assess the extent to which these two statistics, $P(k)$ and $B(k)$, can differentiate between the reionization models based on their respective clustering properties. While all models are compared against one another, the comparison between the MP19 and Fiducial models serves to isolate the effects of spatial inhomogeneities in the ionizing background, as both are calibrated to have the same mean neutral hydrogen fraction $\langle x_{\mathrm{HI}} \rangle$. Since the MP19 model assumes a spatially uniform UV background, it serves as a useful baseline to isolate the effects of spatial inhomogeneities (i.e., patchiness) in the ionization field, although we note that such a comparison is most meaningful with the Fiducial model rather than the more physically motivated patchy scenarios. 
For each model, both $P(k)$ and $B(k)$ were evaluated across multiple sub-volumes to estimate the mean and covariance matrices accurately, using a bootstrap resampling technique with 5000 realizations. This approach enables robust statistical estimation, which is essential for reliable reduced $\chi^2$ analyses.

We then computed the reduced $\chi^2$ statistic between all pairs of models, with the exception of MP19, which was only compared to the Fiducial model. The mean and full covariance matrix for each model were estimated via bootstrap resampling.
 The reduced $\chi^2$ was calculated as
\begin{equation}
\chi^2_{\mathrm{reduced}} = \frac{1}{N_{\mathrm{eff}}} (\bar{\mathbf{x}} - \bar{\mathbf{y}})^T \, \mathbf{C}^{-1} \, (\bar{\mathbf{x}} - \bar{\mathbf{y}}),
\end{equation}
where $\bar{\mathbf{x}}$ and $\bar{\mathbf{y}}$ are the mean vectors corresponding to two reionization models, and $\mathbf{C}$ is the sum of their respective covariance matrices. Here, $N_{\mathrm{eff}}$ represents the effective number of degrees of freedom in the comparison, i.e., the number of independent modes used in the computation. For the power spectrum, which is represented as a 1D vector of 5 $k$-bins, we used $N_{\mathrm{eff}} = 5$. For the bispectrum, we initially flattened the $5 \times 5 \times 9$ array of triangle configurations into a vector of dimension 225. However, due to the high dimensionality and the near-singular nature of the full covariance matrix, we applied Principal Component Analysis (PCA) and retained only the top 30 principal components. Thus, for bispectrum comparisons, we adopted $N_{\mathrm{eff}} = 30$, corresponding to the number of PCA components retained. This dimensionality reduction preserves the dominant modes of variation in the bispectrum signal while ensuring a stable and accurate inversion of the covariance matrix. The choice of $N_{\mathrm{eff}}$ is crucial, as it directly influences the interpretation of the reduced $\chi^2$ values and the associated $p$-values that quantify the statistical distinguishability between models.  

Fig.~\ref{fig:chisq} visually summarizes our findings from this reduced $\chi^2$ analysis. Each point on the plot corresponds to a unique pairwise comparison of the eight reionization models. Points positioned above the diagonal dashed line indicate model pairs for which the bispectrum yields higher reduced $\chi^2$ values than the power spectrum, suggesting enhanced sensitivity to differences in ionization structure. Conversely, points closer to or below the diagonal line represent cases where the two statistics provide comparable levels of discrimination. The shaded region on the left-hand side highlights model pairs with reduced $\chi^2$ values less than unity for the power spectrum, indicating limited separability based on two-point statistics alone, even when the bispectrum shows stronger responses.

From our analysis, it is evident that the bispectrum systematically exhibits enhanced discriminative power compared to the power spectrum. For instance, despite the Fiducial and MP19 models having comparable average neutral fractions, they exhibit marked morphological differences clearly resolved by the bispectrum better than the power spectrum. This superior sensitivity of the bispectrum highlights its effectiveness in capturing non-linearities and higher-order clustering features directly linked to the underlying ionization morphology.

Moreover, subtle differences among models, such as Democratic versus Oligarchic and QSO-assisted versus QSO-only, which are challenging to resolve with the power spectrum, are better distinguished by the bispectrum. Notably, these model pairs appear in the left portion of Fig.\ref{fig:chisq}, where the reduced $\chi^2$ from the power spectrum remains below unity, indicating limited discriminative power, whereas the bispectrum yields a reduced $\chi^2$ greater than 1, signifying successful distinction. This region highlights cases where the power spectrum under the current setup lacks sensitivity. While these differences were not immediately evident from plots of the bispectrum as a function of $k$ alone, incorporating the full set of triangular configurations (see Fig.\ref{Fig:BS_triangleindex}) through the PCA-reduced representation enabled the bispectrum to capture the subtle morphological variations driven by different ionization sources and mechanisms. This result highlights the added value of higher-order statistics, which probe mode couplings beyond the reach of two-point statistics, in capturing the structural complexity of reionization.

These findings align with and extend previous literature investigating the bispectrum statistics in other contexts, such as the 21-cm signal from the EoR \citep{Majumdar2018, Ghara2020}, reinforcing the necessity of incorporating bispectrum measurements into the analyses of future observational surveys. Upcoming large-scale 3D LAE surveys by JWST, MUSE, and next-generation integral-field spectrographs will greatly benefit from the additional morphological insights provided by bispectrum measurements, enabling more precise constraints on the astrophysical processes driving reionization.

\section{Conclusion and Discussion}
\label{Sec:Conclusion}

In this work, we have conducted a comprehensive investigation into the potential of the bispectrum (BS) of high-redshift Ly$\alpha$ emitters (LAEs) as a sensitive and robust probe of the EoR. Using state-of-the-art hydrodynamical simulations from the Sherwood-Relics suite \citep{puchwein2023}, post-processed with the radiative transfer code \atonhe\ \citep{asthana2024}, we generated realistic LAE distributions across a wide range of physically motivated reionization models. These included variations in ionization history (e.g., Fiducial, Extremely Early, and uniform UV background models), as well as differences in ionizing source populations (Democratic vs. Oligarchic) and QSO contributions (QSO-assisted vs. QSO-only).

Our simulations were carried out in a $(160\,h^{-1}\mathrm{cMpc})^3$ box with $2 \times 2048^3$ particles, and incorporated photon propagation from ionizing sources with halo-mass-dependent emissivities and assigned Ly$\alpha$ luminosities using calibrated conditional luminosity functions. Observational effects, including redshift-space distortions and IGM attenuation, were modeled to ensure realism of our modelling.

A key outcome of our study is the better sensitivity of the bispectrum compared to the traditional power spectrum to differences between reionization models. While the power spectrum captures two-point clustering and is sensitive primarily to global quantities such as the mean neutral fraction $\langle x_{\mathrm{HI}} \rangle$, the bispectrum encodes higher-order spatial correlations across all triangular configurations. This makes it uniquely suited to probe the morphology and topology of ionized regions during the EoR, particularly in regimes where the underlying field is highly non-Gaussian.

This distinction becomes particularly relevant when comparing models that share similar global properties but differ in the spatial structure of ionization. For instance, while the power spectrum shows only mild differences between the Fiducial and MP19 models at large scales ($k \lesssim 0.2,h,\mathrm{cMpc}^{-1}$), the bispectrum contains additional information that enables their separation. Similarly, models driven by different source populations—such as Democratic versus Oligarchic, or QSO-assisted versus QSO-only—exhibit differences in ionization morphology that are not easily discernible through two-point statistics. 

To quantitatively assess the separability of models, we employed a reduced $\chi^2$ comparison using bootstrapped covariance matrices derived from multiple sub-volumes. While this particular approach served as a useful metric for model comparison in this study, our broader conclusions do not rely on the use of $\chi^2$ specifically. Rather, the key insight is that the bispectrum encodes significantly more morphological information than the power spectrum, and thus provides a more powerful discriminator across a wide range of physically distinct models. 

{ While our results demonstrate the improved separability of reionization models using the bispectrum, it is important to note that degeneracies with other astrophysical parameters may still persist. For instance, variations in the escape fraction of ionizing photons, redshift evolution of intrinsic galaxy bias, or small-scale feedback processes could potentially imprint features in the LAE bispectrum that resemble those produced by changes in ionization morphology. Although our current setup controls for these parameters to isolate the impact of ionization morphology, future work should systematically explore how variations in such astrophysical prescriptions propagate into clustering observables, and whether the bispectrum can effectively disentangle them. This will be especially crucial when interpreting observational measurements, where these parameters are not independently constrained.}

We also examined the evolution of LAE clustering with redshift, finding that both the power spectrum and bispectrum increase in amplitude at higher redshifts ($z = 7.14$, 6.6, and 5.756), consistent with expectations of stronger small-scale clustering and more inhomogeneous ionization fields in the early phases of reionization. The inclusion of peculiar velocities induces a moderate suppression in both statistics, more noticeably in the bispectrum, highlighting the need for accurate redshift-space distortion modeling when comparing with observational data.

While our focus in this work has been on the power spectrum and bispectrum of LAEs as tools to distinguish between reionization models, we note that other summary statistics, particularly the evolution of the Ly$\alpha$ luminosity function and the equivalent width (EW) distribution, have been widely used to constrain the ionization state of the IGM \citep[e.g.,][]{Konno2018, mason2018, Stark2010}. These observables are sensitive primarily to the integrated visibility of Ly$\alpha$ photons, and can therefore provide constraints on the global neutral hydrogen fraction $\langle x_{\mathrm{HI}} \rangle$ over time. However, such integrated measures lack direct spatial information about the topology of reionization. In contrast, clustering statistics like the power spectrum and especially the bispectrum are sensitive to the spatial structure of ionized and neutral regions, enabling morphological discrimination between models that may produce similar luminosity functions or EW distributions. Thus, our bispectrum-based approach provides complementary information that enhances model separability beyond what is achievable through luminosity or EW statistics alone. A joint analysis combining these different observables will be a powerful direction for future work.

Looking ahead, one of the most promising directions enabled by this work lies in the tomographic use of LAE distributions to constrain reionization morphology. With the advent of large, deep LAE surveys from JWST and future facilities such as the Roman Space Telescope and ELTs, it will become increasingly feasible to map the three-dimensional distribution of LAEs across cosmic volumes. These high-precision redshift surveys open the door to neutral hydrogen tomography using LAEs, in which the large-scale topology of ionized regions can be inferred directly from observed LAE clustering. The bispectrum should play a central role in such tomographic efforts, as it captures non-Gaussian signatures and small-scale morphology with significantly higher sensitivity than traditional clustering analyses.

In addition, cross-correlation studies involving LAEs and the 21-cm signal from experiments such as SKA, HERA, and LOFAR are expected to break degeneracies between astrophysical parameters and reionization models. The bispectrum is well suited to exploit such synergies, as it probes the ionized phase and directly traces the effects of source distribution and feedback on the evolving reionization topology \citep{kubota2016, hutter2020}. Combined analyses leveraging both the 21-cm and LAE bispectrum information should provide orthogonal constraints on the timing, morphology, and drivers of reionization.

Despite these promising results, we note a key limitation arising from the finite size of our simulation volume. The $(160\,h^{-1}\mathrm{cMpc})^3$ box restricts access to the largest-scale modes relevant for cosmic variance and may underestimate the true variance in LAE clustering statistics, particularly on large scales. This limitation also affects the precision of bispectrum estimates, which rely on sampling a wide range of triangle configurations across scales. Future simulations conducted in significantly larger cosmological volumes will be essential for capturing these large-scale modes, improving statistical robustness, and enabling more reliable comparisons with forthcoming wide-field observational surveys.

A further caveat arises when translating these statistics to observed LAE samples. Real-world surveys introduce additional complexities, particularly due to the window function imposed by survey geometry and selection effects. Improper treatment of this window function can result in both inflated uncertainties and systematic biases in the measured power spectrum and bispectrum. While this issue is well recognized for two-point statistics, it is expected to be equally important for the bispectrum due to its sensitivity to large-scale mode coupling. { In practice, this is especially challenging for the bispectrum, as the window function couples multiple modes and triangle configurations in nontrivial ways. Several strategies have been developed to address this. In Fourier space, one may employ pseudo-bispectrum estimators with correction matrices \citep[for example]{Scoccimarro1999,Sugiyama2019}, or forward-model the expected convolved bispectrum by applying the survey mask to theoretical predictions, as done in large-scale structure analyses \citep[for example]{GilMarin2015}.} These observational limitations will ultimately determine how effectively reionization models can be distinguished using LAE clustering. In this context, configuration-space correlation functions may offer complementary insights, as they are somewhat more robust to complex survey geometries. { In this context, configuration-space three-point correlation functions \citep[for example]{SlepianEisenstein2015,SlepianEisenstein2018} provide a complementary approach, as they are more localized and less sensitive to complex survey geometries. Future work should incorporate and test such methods in realistic end-to-end pipelines using mocks that include survey masks, noise, and selection effects, to assess their impact on high-redshift bispectrum constraints.}

In conclusion, our study highlights the bispectrum of LAEs as a powerful higher-order statistic capable of distinguishing reionization models with different ionization morphologies and source models. By leveraging the additional spatial information encoded in three-point correlations, the bispectrum provides a critical extension to two-point statistics and enhances our ability to probe the non-Gaussian features of the EoR. With future observational data and simulation capabilities on the horizon, bispectrum-based analyses, particularly in tomographic settings, are poised to play a central role in unraveling the physics of cosmic reionization and the nature of early galaxy formation.

\section*{Acknowledgements}
It is a pleasure to thank Jason Prochaska for the insightful discussions from which the idea for this project emerged.
This work was performed partially using the DiRAC Data Intensive service CSD3 at the University of Cambridge, managed by the University of Cambridge University Information Services on behalf of the STFC DiRAC HPC Facility, and the DiRAC Extreme Scaling service Tursa at the University of Edinburgh, managed by the Edinburgh Parallel Computing Centre on behalf of the STFC DiRAC HPC Facility (www.dirac.ac.uk). The DiRAC components of CSD3 at Cambridge and Tursa at Edinburgh were funded by BEIS, UKRI and STFC capital funding and STFC operations grants. DiRAC is part of the UKRI Digital Research Infrastructure. The project was also supported by a Swiss National Supercomputing Centre (CSCS) grant under project ID s1114. The authors acknowledge the computational resources provided by the Department of Theoretical Physics, Tata Institute of Fundamental Research (TIFR). SA also thanks the Science and Technology Facilities Council for a PhD studentship (STFC grant reference ST/W507362/1) and the University of Cambridge for providing a UKRI International Fees Bursary. The work has been performed as part of the DAE-STFC collaboration `Building Indo-UK collaborations towards the Square Kilometre Array' (STFC grant reference ST/Y004191/1). JSB acknowledges support by STFC consolidated grant ST/X000982/1. GK also gratefully acknowledges the KICC Medium-Term Visitor programme. This research used resources of the Oak Ridge Leadership Computing Facility at the Oak Ridge National Laboratory, which is supported by the Office of Science of the U.S. Department of Energy under Contract No. DE-AC05-00OR22725. These resources were granted via INCITE AST206. We also acknowledge the use of OpenAI's ChatGPT for support with code implementation, text refinement, and certain analytical calculations. 


\section*{Data Availability}

The data and code underlying this article will be shared upon reasonable request to the corresponding author. We make our LAE model publicly available as the code \textsc{SiMPLE-Gen}, which can be accessed at \url{https://github.com/soumak-maitra/SiMPLE-Gen}.



\bibliographystyle{mnras}
\bibliography{main} 



\clearpage
\newpage
\appendix

\section{Shot Noise Contribution to the Power Spectrum and the Bispectrum}
\label{A:Shot_noise}

We consider a set of discrete tracers (in our case galaxies) whose positions are denoted by \(\{\bm{x}_i\}\). The discreteness of the tracers introduces \emph{shot noise} corrections to the power spectrum and bispectrum.
The number density of the tracers in real space is given as
\begin{equation}
\label{eq:C1}
n(\bm{x}) = \sum_{i} \delta_{\rm D}\left(\bm{x} - \bm{x}_i\right),
\end{equation}
where \(\delta_D\) is the Dirac delta function. Its ensemble average gives the mean density, 
\begin{equation}
\label{eq:C2}
\bar{n} = \langle n(\bm{x}) \rangle.
\end{equation}
The overdensity of the tracer number density is given by
\begin{equation}
\label{eq:C3}
\delta(\bm{x}) = \frac{n(\bm{x}) - \bar{n}}{\bar{n}},
\end{equation}
which satisfies \(\langle \delta(\bm{x}) \rangle = 0\).

\subsection{Real-Space Clustering Statistics}

\subsubsection{Two-Point Correlation Function}

The real-space two-point correlation function of the discrete tracer is defined as
\begin{equation}
\label{eq:C4}
\xi(\bm{r}) = \langle \delta(\bm{x}) \, \delta(\bm{x}+\bm{r}) \rangle = \frac{1}{\bar{n}^2} \left\langle n(\bm{x})n(\bm{x+r}) \right\rangle - 1.
\end{equation}
Using Equation~(\ref{eq:C1}), the ensemble average of the product of number densities at $\bm{x}$ and $\bm{x}'$ can be expressed as
\begin{equation}
\label{eq:C5}
\langle n(\bm{x}) \, n(\bm{x}') \rangle 
= \left\langle \sum_{i,j} \delta_D(\bm{x} - \bm{x}_i)\, \delta_D(\bm{x}' - \bm{x}_j) \right\rangle.
\end{equation}
In this summation there are two contributions:
\begin{enumerate}
    \item The \emph{diagonal} term with \(i=j\), which gives
    \begin{equation}
    \label{eq:C6}
    \begin{aligned}
    \langle n(\bm{x}) \, n(\bm{x}') \rangle_{i=j} 
    &= \sum_i \langle \delta_{\rm D}(\bm{x} - \bm{x}_i) \, \delta_{\rm D}(\bm{x}' - \bm{x}_i) \rangle\\
    &= \bar{n}\, \delta_{\rm D}(\bm{x}-\bm{x}').
    \end{aligned}
    \end{equation} 
    \item The \emph{off-diagonal} (or \(i\neq j\)) term, which, using Equation~(\ref{eq:C4}), can be written as:
    \begin{equation}
    \label{eq:C7}
    \langle n(\bm{x}) \, n(\bm{x}') \rangle_{i\neq j} 
    = \bar{n}^2 \left[ 1 + \xi(\bm{x}-\bm{x}') \right].
    \end{equation}
\end{enumerate}
Thus, combining both contributions we have
\begin{equation}
\label{eq:C8}
\langle n(\bm{x}) \, n(\bm{x}') \rangle 
= \bar{n}\, \delta_D(\bm{x}-\bm{x}') + \bar{n}^2 \left[ 1 + \xi(\bm{x}-\bm{x}') \right].
\end{equation}
It is important to note that the delta function term in Equation~(\eqref{eq:C8}) is the manifestation of shot noise in real space.

\subsubsection{Three-Point Correlation Function}

The three-point correlation can be written as
\begin{equation}
\label{eq:C16}
\begin{aligned}
\zeta(\mathbf{x}_1,\mathbf{x}_2,\mathbf{x}_3) &= \langle \delta(\mathbf{x}_1)\delta(\mathbf{x}_2)\delta(\mathbf{x}_3) \rangle \\
&= \frac{\langle n(\mathbf{x}_1)n(\mathbf{x}_2)n(\mathbf{x}_3) \rangle}{\bar{n}^3} - \Big[\frac{\langle n(\mathbf{x}_1)n(\mathbf{x}_2) \rangle}{\bar{n}^2} + 2~\mathrm{cyc.} \Big] + 2
\end{aligned}
\end{equation}
The ensemble average of the product of number densities at $\bm{x}_1, \bm{x}_2$ and $\bm{x}_3$ can be written as
\begin{equation}
\label{eq:C9}
\begin{aligned}
\langle n(\bm{x}_1) \, n(\bm{x}_2) \, n(\bm{x}_3) \rangle 
&= \Bigg\langle \sum_{i,j,k} \delta_{\rm D}(\bm{x}_1-\bm{x}_i)\, \delta_{\rm D}(\bm{x}_2-\bm{x}_j) \notag \\
&\quad \times \delta_{\rm D}(\bm{x}_3-\bm{x}_k) \Bigg\rangle.
\end{aligned}
\end{equation}
The evaluation involves several contributions:
\begin{enumerate}
    \item The fully distinct index contribution (\(i\neq j\neq k\)) representing the three-point correlation function, \(\zeta(\bm{x}_1,\bm{x}_2,\bm{x}_3)\) given as 
    \begin{equation}      
\left< \sum_{i,j,k} \delta_{\rm D}(\mathbf{x}_1 - \mathbf{x}_i) \delta_{\rm D}(\mathbf{x}_2 - \mathbf{x}_j) \delta_{\rm D}(\mathbf{x}_3 - \mathbf{x}_k) \right>.
    \end{equation}
    \item Contributions where two indices coincide (for example, \(i=j\) with \(k\) distinct) yield terms proportional, given as
    \begin{equation}
        \left< \sum_{i,j} \delta_{\rm D}({\bf x}_1 - {\bf x}_i) \delta_{\rm D}({\bf x}_2 - {\bf x}_j) \delta_{\rm D}({\bf x}_3 - {\bf x}_j) \right> + 2 \, \mathrm{cyc.}
    \end{equation}
    \item Finally, the term where all three indices are equal, given as
    \begin{equation}
        \left< \sum_i \delta_{\rm D}(\mathbf{x}_1 - \mathbf{x}_i) \delta_{\rm D}(\mathbf{x}_2 - \mathbf{x}_i) \delta_{\rm D}(\mathbf{x}_3 - \mathbf{x}_i) \right>
    \end{equation}
\end{enumerate}
Adding all the contributions together, the full expression can be written as
\begin{align}
\label{eq:C10}
\langle n(\bm{x}_1)&\, n(\bm{x}_2) \, n(\bm{x}_3) \rangle = \bar{n}^3 \left[ 1 + \xi(\bm{x}_1-\bm{x}_2) + \xi(\bm{x}_2-\bm{x}_3) + \xi(\bm{x}_3-\bm{x}_1) \right] \nonumber \\
&+ \bar{n}^2 \Bigl[ \delta_{\rm D}(\bm{x}_1-\bm{x}_2) (1+\xi(\bm{x}_1-\bm{x}_2) \ + \ 2\mathrm{cyc.} \Bigr] \nonumber \\
&+ \bar{n}\, \delta_{\rm D}(\bm{x}_1-\bm{x}_2)\,\delta_{\rm D}(\bm{x}_2-\bm{x}_3)+ \bar{n}^3\zeta_{\mathrm{intr.}}(\bm{x}_1,\bm{x}_2,\bm{x}_3),
\end{align}
where $\zeta_{\mathrm{intr.}}(\bm{x}_1,\bm{x}_2,\bm{x}_3)$ is the intrinsic three-point correlation.
Using Equation~(\ref{eq:C10}), \ref{eq:C16} and \ref{eq:C4}, one can write,
\begin{equation}
\label{eq:C18}
\begin{aligned}
\zeta(\mathbf{x}_1, \mathbf{x}_2, \mathbf{x}_3) &= \frac{1}{\bar{n}^2} \delta_{\mathrm{D}}(\mathbf{x}_1 - \mathbf{x}_2) \delta_{\mathrm{D}}(\mathbf{x}_2 - \mathbf{x}_3) \\
&\quad + \left[ \frac{\delta_{\mathrm{D}}(\mathbf{x}_2 - \mathbf{x}_3)}{\bar{n}} \xi_{12} + 2 \, \mathrm{cyc.} \right] + \zeta_{\mathrm{intr.}}.
\end{aligned}
\end{equation}

\subsection{Transition to Fourier Space}

In Fourier space, the two-point and three-point functions are expressed in terms of the power spectrum \(P(\bm{k})\) and bispectrum \(B(\bm{k}_1,\bm{k}_2,\bm{k}_3)\), respectively.

\subsubsection{Power Spectrum in Fourier Space}
Fourier transforming the two-point function in Equation~(\eqref{eq:C8}) yields two contributions. The \(\bar{n}^2\) part (associated with clustering) transforms into \(P_{\rm clust}(\bm{k})\), whereas the term containing the delta function transforms straightforwardly:
\begin{equation}
\label{eq:C14}
\int \dd^3 r\, e^{-i\bm{k}\cdot\bm{r}}\, \delta_{\rm D}(\bm{r})
= 1.
\end{equation}
Thus, one obtains
\begin{equation}
\label{eq:C15}
P(\bm{k}) = P_{\rm clust}(\bm{k}) + \frac{1}{\bar{n}},
\end{equation}
where the \(1/\bar{n}\) term is the Fourier-space representation of shot noise.

\subsubsection{Bispectrum in Fourier Space}

Similar to the two-point correlation function, transforming the three-point correlation function from Equation~(\ref{eq:C18}) into Fourier space gives,
\begin{equation}
\begin{aligned}
\label{eq:C17}
B(\bm{k}_1,\bm{k}_2,\bm{k}_3) &= \frac{1}{\bar{n}} \left[ P_{\rm clust}(\bm{k}_1) + P_{\rm clust}(\bm{k}_2) + P_{\rm clust}(\bm{k}_3) \right]\\
&+ \frac{1}{\bar{n}^2} + B_{\rm clust}(\bm{k}_1,\bm{k}_2,\bm{k}_3).
\end{aligned}
\end{equation}
Thus, the shot noise contributions in Fourier space include both a term proportional to \(1/\bar{n}\) (weighted by the power spectrum terms) and a constant term \(1/\bar{n}^2\).

\section{Triangle Indices}
\label{Sec:Triangle_Indices}
In our analysis of the bispectrum, we consider triangular configurations defined by three wavevectors, \(\mathbf{k}_1\), \(\mathbf{k}_2\), and \(\mathbf{k}_3\), satisfying the closure relation \(\mathbf{k}_1 + \mathbf{k}_2 + \mathbf{k}_3 = 0\). These configurations can be uniquely specified by the magnitudes of two sides, \(k_1\), \(k_2\), and the enclosed angle \(\theta\) between these two sides. Throughout this paper, we systematically index these triangles for convenience of visualization and comparison across different reionization models.

\begin{figure*}
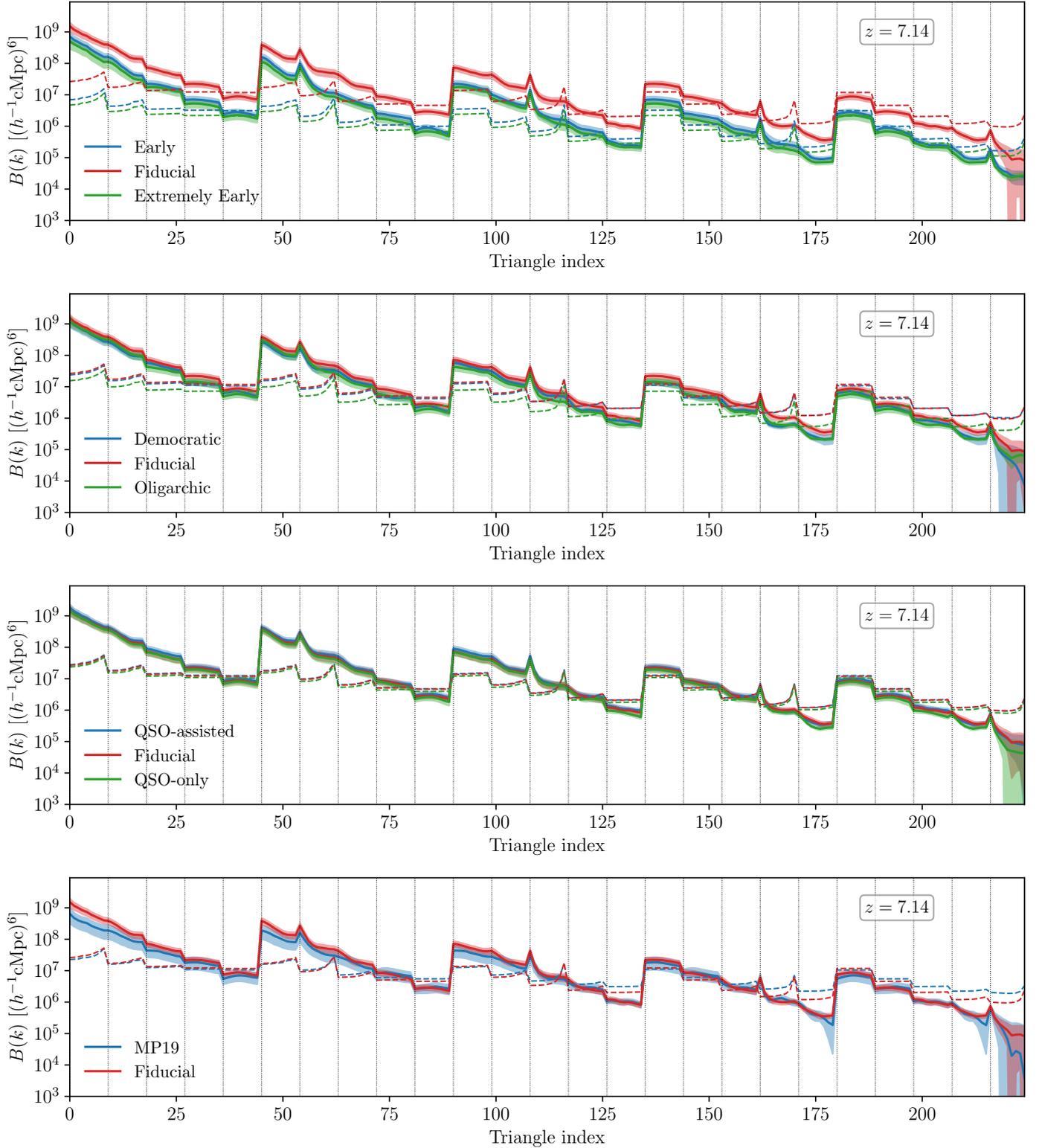


    \includegraphics[width=19cm, trim={0.4cm 8cm 0cm 0.5cm}, clip]{Plots/Bispectrum_ion.pdf}%
    
    \includegraphics[width=19cm, trim={0.4cm 8cm 0cm 0.5cm}, clip]{Plots/Bispectrum_ion2.pdf}

    \includegraphics[width=19cm, trim={0.4cm 8cm 0cm 0.5cm}, clip]{Plots/Bispectrum_ion3.pdf}

    \includegraphics[width=19cm, trim={0.4cm 8cm 0cm 0.5cm}, clip]{Plots/Bispectrum_ion4.pdf}%
	\caption{The 3D bispectrum $B(k)$ of our modeled \lya\ emitters accounting for CGM/IGM attenuation at $z=7.14$ for all triangular configurations, shown as a function of the Triangle Index. The Triangle Index is an integer label used to enumerate all unique triangle configurations in $k$-space, ordered by increasing side lengths and shape type (e.g., squeezed, isosceles, elongated). The top panel compares $B(k)$ for the Early, Fiducial, and Extremely Early reionization models. The second panel contrasts the Democratic, Fiducial, and Oligarchic source models. The third panel compares the QSO-assisted, Fiducial, and QSO-only models. The bottom panel compares the MP19 and Fiducial models. The mean and $68\%$ confidence intervals of $B(k)$ are estimated by bootstrap resampling: 8 sub-volumes are randomly drawn from a total of 27 overlapping sub-volumes of size $(80\,h^{-1}\mathrm{cMpc})^3$. Each bootstrap realization is constructed to represent the full simulation volume $(160\,h^{-1}\mathrm{cMpc})^3$. The dashed curves show the bispectrum shot noise from Equation~\ref{Eq:B_shotnoise}.}
\label{Fig:BS_triangleindex}
\end{figure*}

The indexing procedure adopted is as follows: we bin \(k_1\) and \(k_2\) logarithmically into 5 bins spanning the range \(0.1 \leq k \leq 1.6\,h\,\mathrm{cMpc}^{-1}\), and we bin the angle \(\theta\) uniformly into nine bins spanning the range \(0.13\pi \leq \theta \leq 0.93\pi\). Each unique combination of \(k_1\), \(k_2\), and \(\theta\) corresponds to a single triangle index. The mapping from the triangle indices used in our bispectrum and reduced bispectrum plots to their corresponding \((k_1, k_2, \theta)\) configurations is provided in Table~\ref{tab:triplets}. The bispectrum corresponding to these triangle indices in shown in Fig.~\ref{Fig:BS_triangleindex}.
This indexing allows a direct and systematic comparison of the bispectrum amplitude and shape across different reionization models and observational conditions, facilitating insights into the non-Gaussian signatures and the underlying morphology of reionization captured by different triangle shapes.

\onecolumn
\begin{longtable}{*{12}{c}}
\caption{Values of \((k_{1}, k_{2}, \theta)\) triplets for all triangle indices. All \(k\) values are in units of $h$~cMpc\(^{-1}\) and \(\theta\) is in radians.} \label{tab:triplets} \\
\toprule
Triangle Index & \(k_{1}\) & \(k_{2}\) & \(\theta\) & Triangle Index & \(k_{1}\) & \(k_{2}\) & \(\theta\) & Triangle Index & \(k_{1}\) & \(k_{2}\) & \(\theta\) \\
\midrule
\endfirsthead
\multicolumn{12}{c}%
{{\tablename\ \thetable{} -- continued from previous page}} \\
\toprule
Triangle Index & \(k_{1}\) & \(k_{2}\) & \(\theta\) & Triangle Index & \(k_{1}\) & \(k_{2}\) & \(\theta\) & Triangle Index & \(k_{1}\) & \(k_{2}\) & \(\theta\) \\
\midrule
\endhead
\midrule \multicolumn{12}{r}{{Continued on next page}} \\
\midrule
\endfoot
\bottomrule
\endlastfoot
 1 & 0.1 & 0.1 & $\theta=0.13\pi$ & 2 & 0.1 & 0.1 & $\theta=0.23\pi$ & 3 & 0.1 & 0.1 & $\theta=0.33\pi$ \\
4 & 0.1 & 0.1 & $\theta=0.43\pi$ & 5 & 0.1 & 0.1 & $\theta=0.53\pi$ & 6 & 0.1 & 0.1 & $\theta=0.63\pi$ \\
7 & 0.1 & 0.1 & $\theta=0.73\pi$ & 8 & 0.1 & 0.1 & $\theta=0.83\pi$ & 9 & 0.1 & 0.1 & $\theta=0.93\pi$ \\
10 & 0.1 & 0.2 & $\theta=0.13\pi$ & 11 & 0.1 & 0.2 & $\theta=0.23\pi$ & 12 & 0.1 & 0.2 & $\theta=0.33\pi$ \\
13 & 0.1 & 0.2 & $\theta=0.43\pi$ & 14 & 0.1 & 0.2 & $\theta=0.53\pi$ & 15 & 0.1 & 0.2 & $\theta=0.63\pi$ \\
16 & 0.1 & 0.2 & $\theta=0.73\pi$ & 17 & 0.1 & 0.2 & $\theta=0.83\pi$ & 18 & 0.1 & 0.2 & $\theta=0.93\pi$ \\
19 & 0.1 & 0.4 & $\theta=0.13\pi$ & 20 & 0.1 & 0.4 & $\theta=0.23\pi$ & 21 & 0.1 & 0.4 & $\theta=0.33\pi$ \\
22 & 0.1 & 0.4 & $\theta=0.43\pi$ & 23 & 0.1 & 0.4 & $\theta=0.53\pi$ & 24 & 0.1 & 0.4 & $\theta=0.63\pi$ \\
25 & 0.1 & 0.4 & $\theta=0.73\pi$ & 26 & 0.1 & 0.4 & $\theta=0.83\pi$ & 27 & 0.1 & 0.4 & $\theta=0.93\pi$ \\
28 & 0.1 & 0.8 & $\theta=0.13\pi$ & 29 & 0.1 & 0.8 & $\theta=0.23\pi$ & 30 & 0.1 & 0.8 & $\theta=0.33\pi$ \\
31 & 0.1 & 0.8 & $\theta=0.43\pi$ & 32 & 0.1 & 0.8 & $\theta=0.53\pi$ & 33 & 0.1 & 0.8 & $\theta=0.63\pi$ \\
34 & 0.1 & 0.8 & $\theta=0.73\pi$ & 35 & 0.1 & 0.8 & $\theta=0.83\pi$ & 36 & 0.1 & 0.8 & $\theta=0.93\pi$ \\
37 & 0.1 & 1.6 & $\theta=0.13\pi$ & 38 & 0.1 & 1.6 & $\theta=0.23\pi$ & 39 & 0.1 & 1.6 & $\theta=0.33\pi$ \\
40 & 0.1 & 1.6 & $\theta=0.43\pi$ & 41 & 0.1 & 1.6 & $\theta=0.53\pi$ & 42 & 0.1 & 1.6 & $\theta=0.63\pi$ \\
43 & 0.1 & 1.6 & $\theta=0.73\pi$ & 44 & 0.1 & 1.6 & $\theta=0.83\pi$ & 45 & 0.1 & 1.6 & $\theta=0.93\pi$ \\
46 & 0.2 & 0.1 & $\theta=0.13\pi$ & 47 & 0.2 & 0.1 & $\theta=0.23\pi$ & 48 & 0.2 & 0.1 & $\theta=0.33\pi$ \\
49 & 0.2 & 0.1 & $\theta=0.43\pi$ & 50 & 0.2 & 0.1 & $\theta=0.53\pi$ & 51 & 0.2 & 0.1 & $\theta=0.63\pi$ \\
52 & 0.2 & 0.1 & $\theta=0.73\pi$ & 53 & 0.2 & 0.1 & $\theta=0.83\pi$ & 54 & 0.2 & 0.1 & $\theta=0.93\pi$ \\
55 & 0.2 & 0.2 & $\theta=0.13\pi$ & 56 & 0.2 & 0.2 & $\theta=0.23\pi$ & 57 & 0.2 & 0.2 & $\theta=0.33\pi$ \\
58 & 0.2 & 0.2 & $\theta=0.43\pi$ & 59 & 0.2 & 0.2 & $\theta=0.53\pi$ & 60 & 0.2 & 0.2 & $\theta=0.63\pi$ \\
61 & 0.2 & 0.2 & $\theta=0.73\pi$ & 62 & 0.2 & 0.2 & $\theta=0.83\pi$ & 63 & 0.2 & 0.2 & $\theta=0.93\pi$ \\
64 & 0.2 & 0.4 & $\theta=0.13\pi$ & 65 & 0.2 & 0.4 & $\theta=0.23\pi$ & 66 & 0.2 & 0.4 & $\theta=0.33\pi$ \\
67 & 0.2 & 0.4 & $\theta=0.43\pi$ & 68 & 0.2 & 0.4 & $\theta=0.53\pi$ & 69 & 0.2 & 0.4 & $\theta=0.63\pi$ \\
70 & 0.2 & 0.4 & $\theta=0.73\pi$ & 71 & 0.2 & 0.4 & $\theta=0.83\pi$ & 72 & 0.2 & 0.4 & $\theta=0.93\pi$ \\
73 & 0.2 & 0.8 & $\theta=0.13\pi$ & 74 & 0.2 & 0.8 & $\theta=0.23\pi$ & 75 & 0.2 & 0.8 & $\theta=0.33\pi$ \\
76 & 0.2 & 0.8 & $\theta=0.43\pi$ & 77 & 0.2 & 0.8 & $\theta=0.53\pi$ & 78 & 0.2 & 0.8 & $\theta=0.63\pi$ \\
79 & 0.2 & 0.8 & $\theta=0.73\pi$ & 80 & 0.2 & 0.8 & $\theta=0.83\pi$ & 81 & 0.2 & 0.8 & $\theta=0.93\pi$ \\
82 & 0.2 & 1.6 & $\theta=0.13\pi$ & 83 & 0.2 & 1.6 & $\theta=0.23\pi$ & 84 & 0.2 & 1.6 & $\theta=0.33\pi$ \\
85 & 0.2 & 1.6 & $\theta=0.43\pi$ & 86 & 0.2 & 1.6 & $\theta=0.53\pi$ & 87 & 0.2 & 1.6 & $\theta=0.63\pi$ \\
88 & 0.2 & 1.6 & $\theta=0.73\pi$ & 89 & 0.2 & 1.6 & $\theta=0.83\pi$ & 90 & 0.2 & 1.6 & $\theta=0.93\pi$ \\
91 & 0.4 & 0.1 & $\theta=0.13\pi$ & 92 & 0.4 & 0.1 & $\theta=0.23\pi$ & 93 & 0.4 & 0.1 & $\theta=0.33\pi$ \\
94 & 0.4 & 0.1 & $\theta=0.43\pi$ & 95 & 0.4 & 0.1 & $\theta=0.53\pi$ & 96 & 0.4 & 0.1 & $\theta=0.63\pi$ \\
97 & 0.4 & 0.1 & $\theta=0.73\pi$ & 98 & 0.4 & 0.1 & $\theta=0.83\pi$ & 99 & 0.4 & 0.1 & $\theta=0.93\pi$ \\
100 & 0.4 & 0.2 & $\theta=0.13\pi$ & 101 & 0.4 & 0.2 & $\theta=0.23\pi$ & 102 & 0.4 & 0.2 & $\theta=0.33\pi$ \\
103 & 0.4 & 0.2 & $\theta=0.43\pi$ & 104 & 0.4 & 0.2 & $\theta=0.53\pi$ & 105 & 0.4 & 0.2 & $\theta=0.63\pi$ \\
106 & 0.4 & 0.2 & $\theta=0.73\pi$ & 107 & 0.4 & 0.2 & $\theta=0.83\pi$ & 108 & 0.4 & 0.2 & $\theta=0.93\pi$ \\
109 & 0.4 & 0.4 & $\theta=0.13\pi$ & 110 & 0.4 & 0.4 & $\theta=0.23\pi$ & 111 & 0.4 & 0.4 & $\theta=0.33\pi$ \\
112 & 0.4 & 0.4 & $\theta=0.43\pi$ & 113 & 0.4 & 0.4 & $\theta=0.53\pi$ & 114 & 0.4 & 0.4 & $\theta=0.63\pi$ \\
115 & 0.4 & 0.4 & $\theta=0.73\pi$ & 116 & 0.4 & 0.4 & $\theta=0.83\pi$ & 117 & 0.4 & 0.4 & $\theta=0.93\pi$ \\
118 & 0.4 & 0.8 & $\theta=0.13\pi$ & 119 & 0.4 & 0.8 & $\theta=0.23\pi$ & 120 & 0.4 & 0.8 & $\theta=0.33\pi$ \\
121 & 0.4 & 0.8 & $\theta=0.43\pi$ & 122 & 0.4 & 0.8 & $\theta=0.53\pi$ & 123 & 0.4 & 0.8 & $\theta=0.63\pi$ \\
124 & 0.4 & 0.8 & $\theta=0.73\pi$ & 125 & 0.4 & 0.8 & $\theta=0.83\pi$ & 126 & 0.4 & 0.8 & $\theta=0.93\pi$ \\
127 & 0.4 & 1.6 & $\theta=0.13\pi$ & 128 & 0.4 & 1.6 & $\theta=0.23\pi$ & 129 & 0.4 & 1.6 & $\theta=0.33\pi$ \\
130 & 0.4 & 1.6 & $\theta=0.43\pi$ & 131 & 0.4 & 1.6 & $\theta=0.53\pi$ & 132 & 0.4 & 1.6 & $\theta=0.63\pi$ \\
133 & 0.4 & 1.6 & $\theta=0.73\pi$ & 134 & 0.4 & 1.6 & $\theta=0.83\pi$ & 135 & 0.4 & 1.6 & $\theta=0.93\pi$ \\
136 & 0.8 & 0.1 & $\theta=0.13\pi$ & 137 & 0.8 & 0.1 & $\theta=0.23\pi$ & 138 & 0.8 & 0.1 & $\theta=0.33\pi$ \\
139 & 0.8 & 0.1 & $\theta=0.43\pi$ & 140 & 0.8 & 0.1 & $\theta=0.53\pi$ & 141 & 0.8 & 0.1 & $\theta=0.63\pi$ \\
142 & 0.8 & 0.1 & $\theta=0.73\pi$ & 143 & 0.8 & 0.1 & $\theta=0.83\pi$ & 144 & 0.8 & 0.1 & $\theta=0.93\pi$ \\
145 & 0.8 & 0.2 & $\theta=0.13\pi$ & 146 & 0.8 & 0.2 & $\theta=0.23\pi$ & 147 & 0.8 & 0.2 & $\theta=0.33\pi$ \\
148 & 0.8 & 0.2 & $\theta=0.43\pi$ & 149 & 0.8 & 0.2 & $\theta=0.53\pi$ & 150 & 0.8 & 0.2 & $\theta=0.63\pi$ \\
151 & 0.8 & 0.2 & $\theta=0.73\pi$ & 152 & 0.8 & 0.2 & $\theta=0.83\pi$ & 153 & 0.8 & 0.2 & $\theta=0.93\pi$ \\
154 & 0.8 & 0.4 & $\theta=0.13\pi$ & 155 & 0.8 & 0.4 & $\theta=0.23\pi$ & 156 & 0.8 & 0.4 & $\theta=0.33\pi$ \\
157 & 0.8 & 0.4 & $\theta=0.43\pi$ & 158 & 0.8 & 0.4 & $\theta=0.53\pi$ & 159 & 0.8 & 0.4 & $\theta=0.63\pi$ \\
160 & 0.8 & 0.4 & $\theta=0.73\pi$ & 161 & 0.8 & 0.4 & $\theta=0.83\pi$ & 162 & 0.8 & 0.4 & $\theta=0.93\pi$ \\
163 & 0.8 & 0.8 & $\theta=0.13\pi$ & 164 & 0.8 & 0.8 & $\theta=0.23\pi$ & 165 & 0.8 & 0.8 & $\theta=0.33\pi$ \\
166 & 0.8 & 0.8 & $\theta=0.43\pi$ & 167 & 0.8 & 0.8 & $\theta=0.53\pi$ & 168 & 0.8 & 0.8 & $\theta=0.63\pi$ \\
169 & 0.8 & 0.8 & $\theta=0.73\pi$ & 170 & 0.8 & 0.8 & $\theta=0.83\pi$ & 171 & 0.8 & 0.8 & $\theta=0.93\pi$ \\
172 & 0.8 & 1.6 & $\theta=0.13\pi$ & 173 & 0.8 & 1.6 & $\theta=0.23\pi$ & 174 & 0.8 & 1.6 & $\theta=0.33\pi$ \\
175 & 0.8 & 1.6 & $\theta=0.43\pi$ & 176 & 0.8 & 1.6 & $\theta=0.53\pi$ & 177 & 0.8 & 1.6 & $\theta=0.63\pi$ \\
178 & 0.8 & 1.6 & $\theta=0.73\pi$ & 179 & 0.8 & 1.6 & $\theta=0.83\pi$ & 180 & 0.8 & 1.6 & $\theta=0.93\pi$ \\
181 & 1.6 & 0.1 & $\theta=0.13\pi$ & 182 & 1.6 & 0.1 & $\theta=0.23\pi$ & 183 & 1.6 & 0.1 & $\theta=0.33\pi$ \\
184 & 1.6 & 0.1 & $\theta=0.43\pi$ & 185 & 1.6 & 0.1 & $\theta=0.53\pi$ & 186 & 1.6 & 0.1 & $\theta=0.63\pi$ \\
187 & 1.6 & 0.1 & $\theta=0.73\pi$ & 188 & 1.6 & 0.1 & $\theta=0.83\pi$ & 189 & 1.6 & 0.1 & $\theta=0.93\pi$ \\
190 & 1.6 & 0.2 & $\theta=0.13\pi$ & 191 & 1.6 & 0.2 & $\theta=0.23\pi$ & 192 & 1.6 & 0.2 & $\theta=0.33\pi$ \\
193 & 1.6 & 0.2 & $\theta=0.43\pi$ & 194 & 1.6 & 0.2 & $\theta=0.53\pi$ & 195 & 1.6 & 0.2 & $\theta=0.63\pi$ \\
196 & 1.6 & 0.2 & $\theta=0.73\pi$ & 197 & 1.6 & 0.2 & $\theta=0.83\pi$ & 198 & 1.6 & 0.2 & $\theta=0.93\pi$ \\
199 & 1.6 & 0.4 & $\theta=0.13\pi$ & 200 & 1.6 & 0.4 & $\theta=0.23\pi$ & 201 & 1.6 & 0.4 & $\theta=0.33\pi$ \\
202 & 1.6 & 0.4 & $\theta=0.43\pi$ & 203 & 1.6 & 0.4 & $\theta=0.53\pi$ & 204 & 1.6 & 0.4 & $\theta=0.63\pi$ \\
205 & 1.6 & 0.4 & $\theta=0.73\pi$ & 206 & 1.6 & 0.4 & $\theta=0.83\pi$ & 207 & 1.6 & 0.4 & $\theta=0.93\pi$ \\
208 & 1.6 & 0.8 & $\theta=0.13\pi$ & 209 & 1.6 & 0.8 & $\theta=0.23\pi$ & 210 & 1.6 & 0.8 & $\theta=0.33\pi$ \\
211 & 1.6 & 0.8 & $\theta=0.43\pi$ & 212 & 1.6 & 0.8 & $\theta=0.53\pi$ & 213 & 1.6 & 0.8 & $\theta=0.63\pi$ \\
214 & 1.6 & 0.8 & $\theta=0.73\pi$ & 215 & 1.6 & 0.8 & $\theta=0.83\pi$ & 216 & 1.6 & 0.8 & $\theta=0.93\pi$ \\
217 & 1.6 & 1.6 & $\theta=0.13\pi$ & 218 & 1.6 & 1.6 & $\theta=0.23\pi$ & 219 & 1.6 & 1.6 & $\theta=0.33\pi$ \\
220 & 1.6 & 1.6 & $\theta=0.43\pi$ & 221 & 1.6 & 1.6 & $\theta=0.53\pi$ & 222 & 1.6 & 1.6 & $\theta=0.63\pi$ \\
223 & 1.6 & 1.6 & $\theta=0.73\pi$ & 224 & 1.6 & 1.6 & $\theta=0.83\pi$ & 225 & 1.6 & 1.6 & $\theta=0.93\pi$ \\
\end{longtable}


\bsp	
\label{lastpage}
\end{document}